\documentclass[aps,pre,twocolumn,showpacs,normalem]{revtex4-1}

\pdfoutput=1

\usepackage{amsmath,amssymb}
\usepackage{bm}
\usepackage{graphicx}

\usepackage{color}

\usepackage{ulem}
\newcommand{\rev}[1]{{#1}}

\begin{document}

\title{
Velocity and displacement statistics in a stochastic model of nonlinear friction showing bounded particle speed
}

\author{Andreas M.\ Menzel}
\email{menzel@thphy.uni-duesseldorf.de}
\affiliation{Institut f\"ur Theoretische Physik II: Weiche Materie, Heinrich-Heine-Universit\"at D\"usseldorf, D-40225 D\"usseldorf, Germany
}

\date{\today}

\begin{abstract}
Diffusion of colloidal particles in a complex environment such as polymer networks or biological cells is a topic of high complexity with significant biological and medical relevance. In such situations, the 
interaction between the surroundings and the particle motion has to be taken into account. We analyze a simplified diffusion model that includes some aspects of a complex 
environment in the framework of a nonlinear friction process: at low particle speeds, friction grows linearly with the particle velocity as for regular viscous friction; it grows more than linearly at higher particle velocities; finally, at a maximum of the possible particle speed the friction diverges. In addition to bare diffusion, we study the influence of a constant drift force acting on the diffusing particle. While the corresponding stationary velocity distributions can be derived analytically, the displacement statistics generally must be determined numerically. However, as a benefit of our model, analytical progress can be made in one case of a special maximum particle speed. The effect of a drift force in this case is analytically determined by perturbation theory. It will be interesting in the future to compare our results to real experimental systems. One realization 
could be magnetic colloidal particles diffusing through a shear-thickening environment such as starch suspensions, possibly exposed to an external magnetic field gradient. 
\end{abstract}

\pacs{82.70.Dd,05.10.Gg,83.60.Fg,66.10.cg}


\maketitle


\section{Introduction} \label{sec:Introduction}

In a complex environment \cite{wang2009anomalous,wang2012brownian,hofling2013anomalous,guan2014even, kalathi2014nanoparticle,babaye2014mobility}, diffusion of a colloidal particle shows properties different from those in typical simple fluid surroundings. The latter case, i.e.\ regular diffusion of non-interacting Brownian particles in a Newtonian fluid, represents a classical textbook example \cite{kubo1991statistical,zwanzig2001nonequilibrium, kampen2007stochastic,gardiner2009stochastic} with well-known characteristics: a mean squared displacement that grows linearly in time as well as a Gaussian velocity distribution and a Gaussian displacement statistics. Interestingly, for diffusion in complex surroundings, e.g.\ on membrane surfaces \cite{wang2009anomalous}, in filament suspensions \cite{wang2009anomalous}, or in a colloidal matrix environment \cite{guan2014even}, some of these features can persist, particularly the linear growth of the mean squared displacement with time. Then, the influence of the surroundings is reflected by deviations of the higher-order displacement moments, resulting in a non-Gaussian displacement statistics \cite{wang2009anomalous,guan2014even,chubynsky2014diffusing}. In these cases, particle motion is driven by the stochastic forces exerted by the thermally fluctuating environment. 

Non-Gaussian displacement statistics were observed in several further contexts, for instance in glass-like states \cite{kegel2000direct}, for actively driven particles \cite{zheng2013non}, or for particles exposed to nonlinear friction. 
One prominent class of nonlinear friction models was used to characterize the stochastic motion of rigid particles on vibrating rigid surfaces \cite{goohpattader2009experimental,goohpattader2010diffusive,goohpattader2011stochastic}. In contrast to the well-known \textit{viscous friction} which increases \textit{linearly} with the speed of the moving object, the so-called \textit{dry friction} between rigid objects is often modeled by a contribution of the Coulomb type \cite{persson2000sliding,kawarada2004non,gennes2005brownian,hayakawa2005langevin, goohpattader2009experimental,goohpattader2010diffusive, touchette2010brownian,goohpattader2011stochastic, baule2011stick,menzel2011effect,talbot2012effect,baule2012singular, baule2013rectification,pototsky2013periodically, kanazawa2015asymptotic,sano2014roles,chen2014large,menzel2015tuned}. This frictional contribution assumes \textit{constant} decelerating forces, independently of the actual speed of the moving object. The corresponding displacement statistics under stochastic motion was analyzed and showed markedly non-Gaussian behavior \cite{goohpattader2009experimental,goohpattader2010diffusive, goohpattader2011stochastic, menzel2011effect,chen2014large}, despite a clear linear increase of the mean squared displacement with proceeding time \cite{menzel2011effect,chen2014large}. 

Here, we likewise address the influence of a \textit{nonlinear} frictional contribution. When compared to the just described dry friction, our nonlinear friction force shows qualitatively opposite behavior. Yet, it may reflect several properties of the motion in a complex 
environment. 

In the limit of small particle speeds, our friction force is linear as for regular viscous friction. This is in agreement with a viscous low-frequency response of viscoelastic surroundings. At higher velocities, our frictional force increases more than linearly and finally diverges. Thus there is a maximum speed by which a particle can be dragged through its environment. Qualitatively, this reflects aspects of the response of a 
complex environment that at high frequencies does not allow viscous motion any more; instead it reacts by a counteracting force that bounds the maximum attainable speed under a certain driving force. 
Such characteristics reflect features of a shear-thickening environment \cite{fall2008shear}. 
In contrast to that, the previously investigated models of Coulomb friction \cite{persson2000sliding,kawarada2004non,gennes2005brownian,hayakawa2005langevin, goohpattader2009experimental,goohpattader2010diffusive, touchette2010brownian,goohpattader2011stochastic, baule2011stick,menzel2011effect,talbot2012effect,baule2012singular, baule2013rectification,pototsky2013periodically, kanazawa2015asymptotic,sano2014roles,chen2014large,menzel2015tuned} could be interpreted as an extreme case of shear-thinning environments. 

What the simplified nonlinear friction model presented here does not cover is the reversible character of the elastic part of a viscoelastic environmental response. That is, if an external force that pulls a particle is released sufficiently quickly, the squeezed environment will to the extent of its remaining stored elastic deformation partially push back the particle. Memory terms \cite{yamamoto2014microscopic} 
would be necessary to include this effect and could be added in a later expansion of our model. 
Thus, strictly speaking, our present formulation of the model applies to the situation of a complex fluid environment, the viscosity of which increases with the impact speed 
and diverges at a certain maximum speed. 

One example that motivates our study and where at least part of the above ingredients might possibly apply stems from drug targeting for cancer treatment \cite{torchilin2000drug,peer2007nanocarriers,bertrand2014cancer}. To direct oncological drugs in the human body to the place of application, i.e.\ to cancer tissue, magnetic colloidal carrier particles were loaded with the medical substance \cite{alexiou2006targeting,tietze2013efficient,zaloga2014development}. It was then demonstrated that strong externally applied magnetic field gradients can be used to direct an elevated fraction of the loaded carrier particles to the target area \cite{dobson2006magnetic,alexiou2006targeting,tietze2013efficient,matuszak2015endothelial}. Our phenomenological model 
might reflect some features of the process of spatial spreading of magnetic colloidal particles in parts of human tissue under a constant external magnetic field gradient. 

We will proceed in the following way. In Sec.~\ref{sec_model}
we introduce the basic equations of our statistical nonlinear friction model and determine the stationary velocity distribution. 
Next, in Sec.~\ref{sec_timeevolution}, we analyze the time evolution of the spatial particle distribution and illustrate the influence of a constant drift force. Such a force could for instance be generated by a constant magnetic field gradient acting on dipolar magnetic colloidal particles. After that, in Sec.~\ref{sec_analytical}, we present an analytical treatment of the problem, where progress can be made for a specific value of the maximum diffusion velocity of the particle. The influence of a constant drift force can be included by perturbational analysis. We 
summarize our results and conclude in Sec.~\ref{sec_conclusions}. 
\rev{In a first Appendix, we briefly contrast results from our nonlinear friction model to those obtained for linear (viscous) friction. Finally, in a second Appendix, some remarks on the influence of a confining harmonic potential are added.}

\section{Nonlinear statistical friction model}\label{sec_model}  

As described in the Introduction, we are interested in the statistics of the motion of a Brownian particle subjected to a nonlinear friction force that may result from a specific environment. For simplicity, we confine our investigation of the particle diffusion to one spatial direction only. Our expression for the friction force acting on the diffusing particle then reads 
\begin{equation}\label{eq_friction}
F_{\mathrm{fr}}(v) = -A\,\tan\left(\frac{\pi v}{2a}\right). 
\end{equation}
Here, $v$ denotes the velocity. The constant $A>0$ sets the strength of the friction force, while the constant $a>0$ sets the maximum speed that the particle can attain. 

This expression for $F_{\mathrm{fr}}$ shows the requested features. 
On the one hand, at low particle speeds $|v|/a\ll1$ it is linear in the velocity $v$ as for regular viscous friction, 
\begin{equation}\label{eq_linearized_friction}
F_{\mathrm{fr}}(v) \approx -\,\frac{\pi A}{2a}\, v \qquad\mbox{ for }\quad\frac{|v|}{a}\ll1. 
\end{equation}
Following Stokes' law for a spherical particle of radius $r$ \cite{dhont1996introduction}, the constant prefactor of $-v$ in this expression sets the friction coefficient and the constants $A$ and $a$ can be related to an effective viscosity $\eta$ of the surrounding medium via 
\begin{equation}
\eta = \frac{1}{12\,r}\frac{A}{a}. 
\end{equation}
On the other hand, the expression in Eq.~(\ref{eq_friction}) diverges for $|v|\rightarrow a$. Thus, the constant $a$ sets the maximum speed that the particle can attain during its diffusive motion through the complex environment. Between these two limits, the friction force monotonously and stronger than linearly increases with increasing particle speed. 

In the present work, we only consider the dilute regime of noninteracting Brownian particles. As a consequence, the equations of motion for one such particle become
\begin{eqnarray}
m\,\frac{dv}{dt} &=& -A\,\tan\left(\frac{\pi v}{2a}\right) +\gamma(t) +M, 
\label{langevin_v}\\
\frac{dx}{dt} &=& v.
\label{langevin_x}
\end{eqnarray}
Here, $m$ is the particle mass, $t$ denotes time, and $x$ gives the spatial position. The stochastic force $\gamma(t)$ is assumed as a Gaussian white process of zero mean, $\langle\gamma(t)\rangle=0$, and correlation $\langle\gamma(t)\gamma(t')\rangle=2Kk_BT\delta(t-t')$, where the strength $K$ in equilibrium is given by a fluctuation-dissipation relation, $k_B$ denotes the Boltzmann constant, $T$ temperature, and $\delta(t-t')$ the delta function. Finally, $M$ is a constant drift force that may for instance be realized by an external magnetic field gradient acting on a dipolar magnetic particle. 

Next, we transform the Langevin Eqs.~(\ref{langevin_v}) and (\ref{langevin_x}) by the standard means \cite{kubo1991statistical,zwanzig2001nonequilibrium, kampen2007stochastic,gardiner2009stochastic,risken1996fokker} to the continuum probability picture in the framework of a Fokker-Planck equation. We obtain for the probability density $f(x,v,t)$ to find at a certain time $t$ a particle with velocity $v$ at position $x$: 
\begin{equation}\label{eq_fp_unscaled}
\partial_t f = -v\,\partial_x f + \partial_v \left[ \frac{A}{m}\,\tan\left(\frac{\pi v}{2a}\right) - \frac{M}{m} \right] f + \frac{Kk_BT}{m^2} \,\partial_v^2 f, 
\end{equation}
together with the normalization condition $\int_{-\infty}^{\infty} \mathrm{d}x\,\int_{-a}^a \mathrm{d}v\,f(x,v,t)=1$. The equation is rescaled by the following replacements: $t=(Kk_BT/A^2)\,t'$, $x=[(Kk_BT)^2/mA^3]\,x'$, $v=(Kk_BT/mA)\,v'$, thus $f=[m^2A^4/(Kk_BT)^3]\,f'$ to keep the normalization condition, $a=(Kk_BT/mA)\,a'$, and $M=A\,M'$. Omitting from now on the primes, our final stochastic model equation becomes
\begin{equation}\label{eq_fp}
\partial_t f = {}-v\,\partial_x f + \partial_v \tan\left(\frac{\pi v}{2a}\right)f - M \,\partial_v f + \partial_v^2 f. 
\end{equation}

From this equation, we find the stationary velocity distribution
\begin{equation}\label{eq_fst}
f_{\mathrm{st}}(v) = \frac{1}{Z}\,\exp\left\{Mv+\frac{2a}{\pi}\ln\left|\cos\left(\frac{\pi v}{2a}\right)\right|\right\}, 
\end{equation}
where the prefactor $1/Z$ follows from the normalization condition $\int_{-a}^a \mathrm{d}v\, f_{\mathrm{st}}(v)=1$. [Scaling back to the initial units, the strength $K$ associated with the stochastic force in equilibrium can be obtained from the requirement of equipartition $m\langle v^2\rangle_{\mathrm{st}}=k_BT$ in the absence of a drift force, i.e.\ for $M=0$.]

\section{Time evolution of the spatial particle distribution}\label{sec_timeevolution}

We now investigate how the statistical spatial particle distribution evolves over time. As indicated above, we here consider a system of non-interacting (dilute) particles. In this sense, our approach coincides with the analysis of the displacement statistics of a single isolated particle. 

Without loss of generality, we may presume that at an initial time $t=0$ a particle is found with high spatial accuracy at position $x=0$. To reflect this positional certainty we assume an initial spatial localization in the form of a sharp Gaussian peak. Furthermore, we may assume that the only action taken at $t=0$ 
is to start our observation. Thus the velocity statistics at $t=0$ is determined by its steady-state distribution. Along these lines, we set for the initial probability distribution 
\begin{equation}\label{eq_init}
f(x,v,t=0) = \frac{1}{\sqrt{\pi}\sigma}\,\exp\left\{-\left(\frac{x}{\sigma}\right)^2\right\}\,f_{\mathrm{st}}(v),
\end{equation}
where we choose $\sigma=0.1$ in our case. Starting from this initial condition, we numerically integrated Eq.~(\ref{eq_fp}) forward in time using a finite difference scheme. Comparable results were obtained by a simple Euler and a fourth order Runge-Kutta method \cite{press1992numerical}. An upwind scheme was employed to discretize the first-order spatial derivatives. The calculation was performed on a regular two-dimensional rectangular grid to cover the $x$-$v$ space. In $v$ direction, this grid is bordered by $v=\pm a$, where we used no-flux boundary conditions to avoid leakage of the probability density. 

Some of our results were tested by direct particle-based simulations. For this purpose, we rescaled Eqs.~(\ref{langevin_v}) and (\ref{langevin_x}) in the same way as described between Eqs.~(\ref{eq_fp_unscaled}) and (\ref{eq_fp}). We consider $N=10^6$ identical non-interacting particles initially located at positions $x(t=0)$ with velocities $v(t=0)$ statistically distributed according to Eq.~(\ref{eq_init}). 
Then we iterate the particle positions and velocities forward until we reach a requested time $t$. At that time, we determine the velocity and spatial distribution functions in the form of discretized histograms.

\subsection{Bare diffusion under nonlinear friction}\label{barediffnonlin}

\begin{figure}
\centerline{\includegraphics[width=7.5cm]{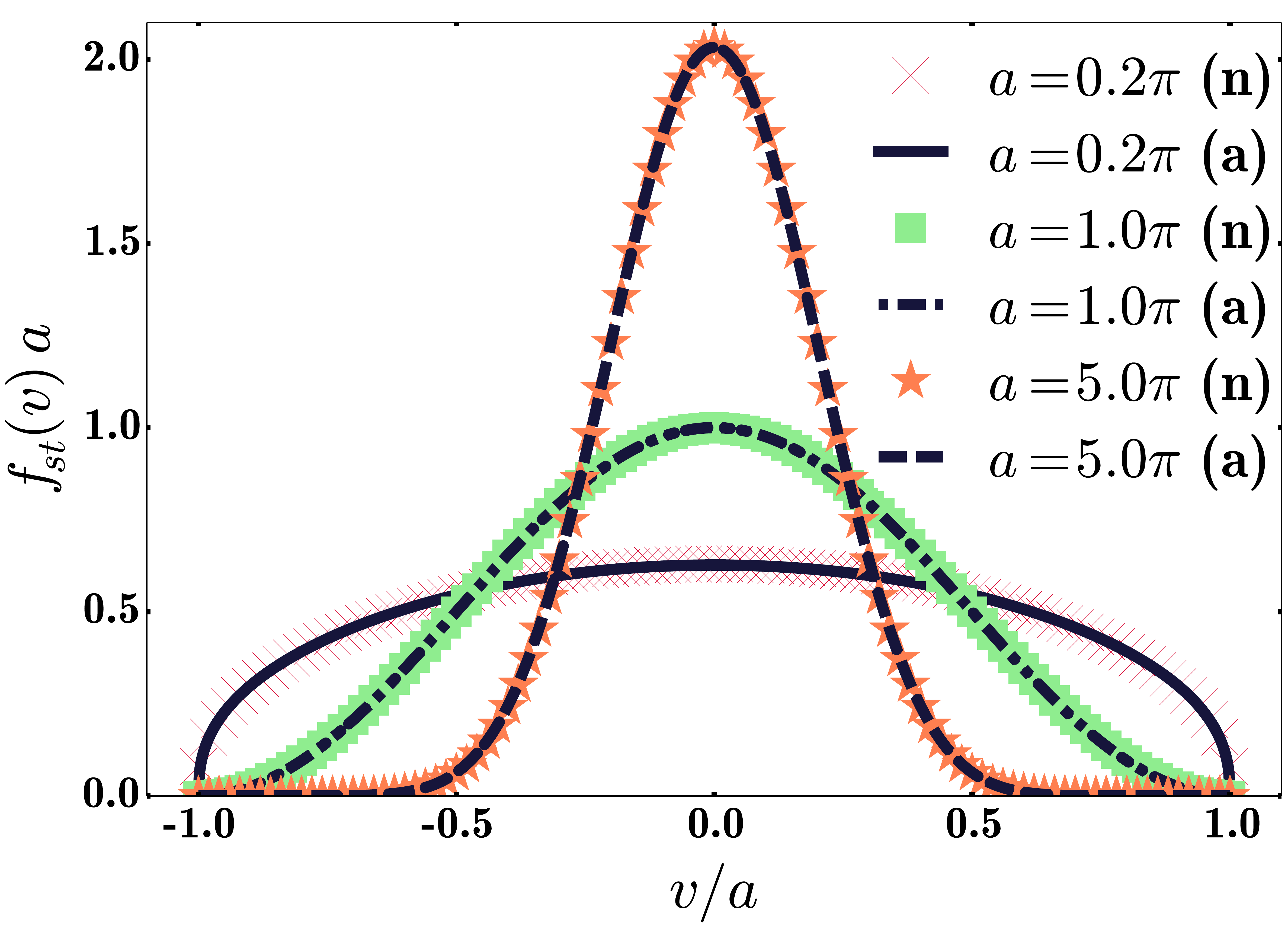}}
\caption{\rev{(Color online)} Comparison of $f_{\mathrm{st}}(v)\equiv f(v,t)$ as obtained via numerical solution of Eqs.~(\ref{eq_fp}) and (\ref{eq_init}) as well as analytically from Eq.~(\ref{eq_fst}). Numerical results are marked by ``(n)'', analytical ones by ``(a)''. The drift force vanishes $M=0$ and different magnitudes of the maximum speed $a$ are considered. In this example, the time to determine $f_{\mathrm{st}}(v)\equiv f(v,t)$ was selected as $t=5$, but is irrelevant. Numerical and analytical results show good agreement.}
\label{fig_fst_Meq0}
\end{figure}

First, we consider the case of vanishing drift force $M=0$. We can test our numerical iteration scheme by integrating our numerical result for $f(x,v,t)$ at a certain time $t$, obtained by forward-iteration of Eq.~(\ref{eq_fp}), over $x$. Analytically, together with our initial condition Eq.~(\ref{eq_init}), we find
\begin{equation}\label{eq_fvt}
f(v,t) := \int_{-\infty}^{\infty}\mathrm{d}x\, f(x,v,t) \equiv f_{\mathrm{st}}(v)
\end{equation}
at all times, because Eq.~(\ref{eq_fp}) becomes $\partial_tf(v,t)\equiv0$ upon integration over $x$ and inserting $f_{\mathrm{st}}(v)$. Numerically obtained example data for $f(v,t)$ are depicted in Fig.~\ref{fig_fst_Meq0} and show good agreement with the analytical results $f_{\mathrm{st}}(v)$ from Eq.~(\ref{eq_fst}). 

\begin{figure}
\centerline{\includegraphics[width=7.5cm]{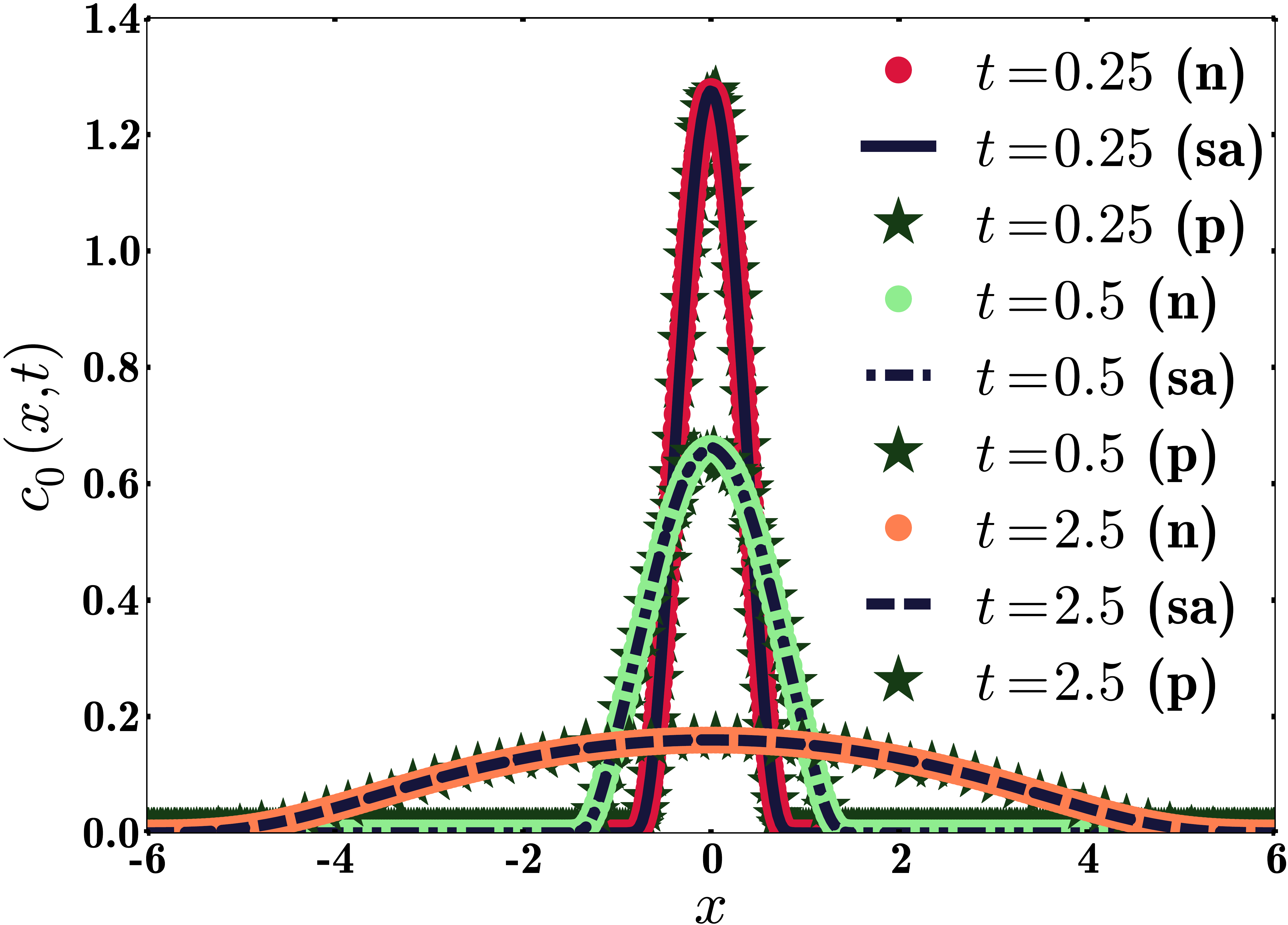}}
\caption{\rev{(Color online)} Time evolution of the spatial distribution function $c_0(x,t)$ for vanishing drift force $M=0$ and maximum speed $a=\pi$ starting from Eq.~(\ref{eq_init}). Results obtained from direct numerical integration of the Fokker-Planck equation Eq.~(\ref{eq_fp}), via the semi-analytical approach presented in Sec.~\ref{sec_analytical}, and from particle-based simulations, marked by ``(n)'', ``(sa)'', and ``(p)'', respectively, show good agreement. [Results from the particle-based simulations are all indicated by the same symbol but are readily identified by the ``(n)''- and ``(sa)''-curves behind which they hide.]}
\label{fig_timeseries_Meq0}
\end{figure}

In analogy to that, we find the time-dependent spatial distribution function $c_0(x,t)$ by integrating out the velocity $v$, 
\begin{equation}\label{eq_c0}
c_0(x,t) := \int_{-a}^{a}\mathrm{d}v\, f(x,v,t).
\end{equation}
The notation $c_0(x,t)$ will become evident in Sec.~\ref{sec_analytical}. Fig.~\ref{fig_timeseries_Meq0} shows the time evolution of $c_0(x,t)$ for one example case. 
As expected, the sharp initial density peak described by Eq.~(\ref{eq_init}) broadens and flattens over time. We tested our results obtained from direct numerical forward-iteration of the Fokker-Planck equation Eq.~(\ref{eq_fp}) by particle-based simulations as explained above. Both routes are in good agreement with each other. Since there is no drift force $M=0$ in Fig.~\ref{fig_timeseries_Meq0}, the distribution curves remain symmetric with respect to $x=0$. That is, odd moments of $c_0(x,t)$ vanish. In particular, $\langle x\rangle(t)=0$ at all times. 

\begin{figure}
\centerline{\includegraphics[width=7.5cm]{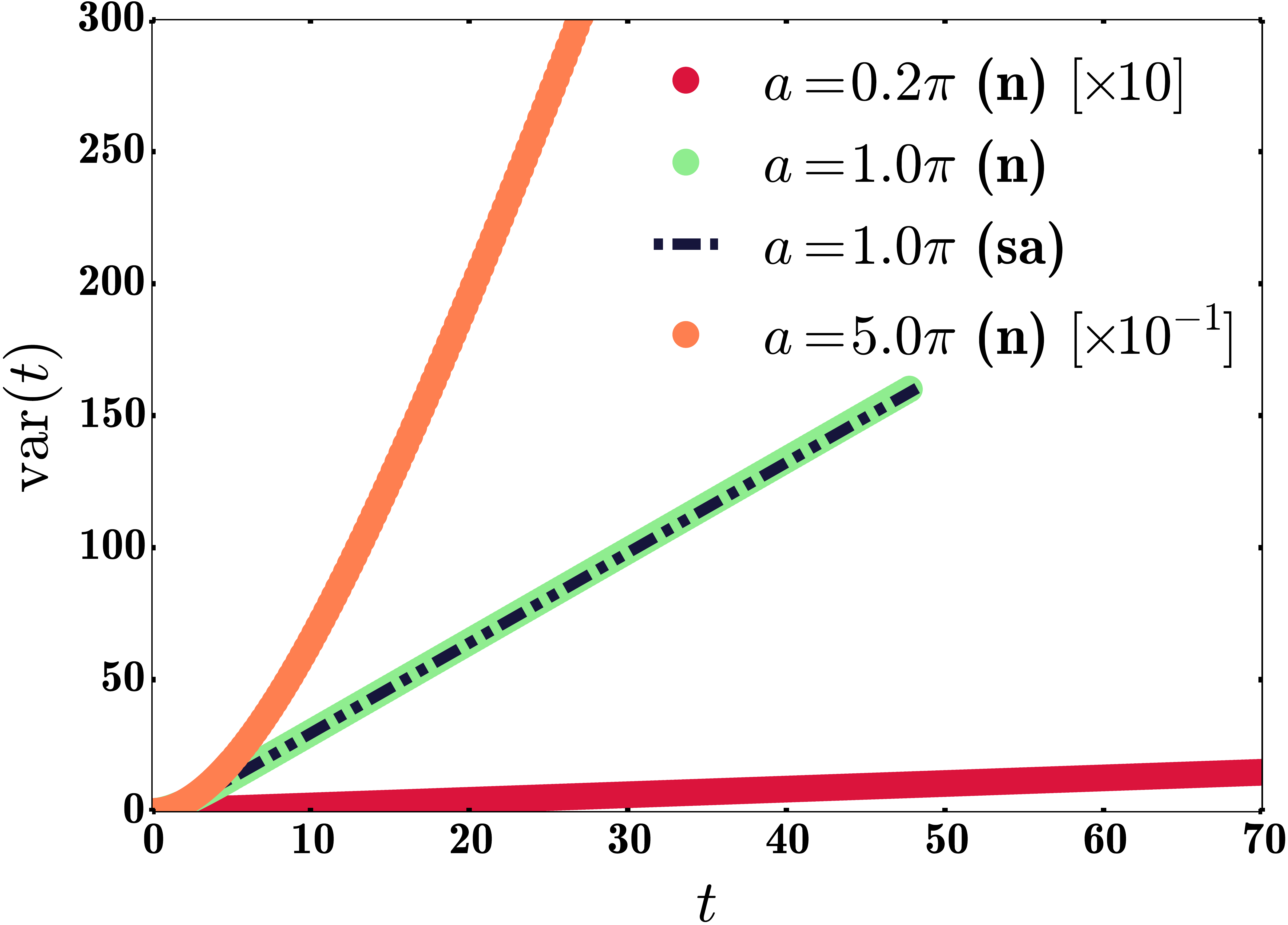}}
\caption{\rev{(Color online)} Variance $\mathrm{var}(t)$ as a function of time $t$ for vanishing drift force $M=0$ and different maximum speeds $a$. We derived the results by numerical forward-iteration of Eqs.~(\ref{eq_fp}) and (\ref{eq_init}), indicated by ``(n)''. In the case of $a=\pi$, the results additionally follow from semi-analytical calculations ``(sa)'' in Sec.~\ref{sec_analytical} via Eqs.~(\ref{eq_c2mp1})--(\ref{eq_e04}). There is good agreement between the curves obtained in the two different ways. The variance appears to grow linearly in time after an initial transient has decayed. Since $\langle x\rangle(t)=0$, the variance here coincides with the mean squared displacement. Data for $a=0.2\pi$ and $a=5.0\pi$ are multiplied by factors $10$ and $10^{-1}$, respectively, for better visualization.}
\label{fig_var}
\end{figure}

When we plot the variance 
\begin{equation}\label{eq_var}
\mathrm{var} = \left\langle \big( x - \langle x\rangle \big)^2 \right\rangle, 
\end{equation}
see Fig.~\ref{fig_var}, we observe a linear increase with time, after an initial transient has decayed. 
Since $\langle x\rangle(t)=0$ at all times, the variance here coincides with the mean squared displacement. Thus, as for regular diffusive processes under linear friction and after the initial transient has decayed, the mean squared displacement here increases linearly with time, which is often still referred to as Brownian diffusion. We observe that the slope of $\mathrm{var}(t)$ is much smaller for lower values of $a$. That is, the spatial distribution function broadens significantly less rapidly. This is in accordance with the more bounded particle velocities at lower values of $a$, reducing the maximum possible particle speed. As a consequence, the spatial probability distribution remains more localized. 

\begin{figure}
\centerline{\includegraphics[width=7.5cm]{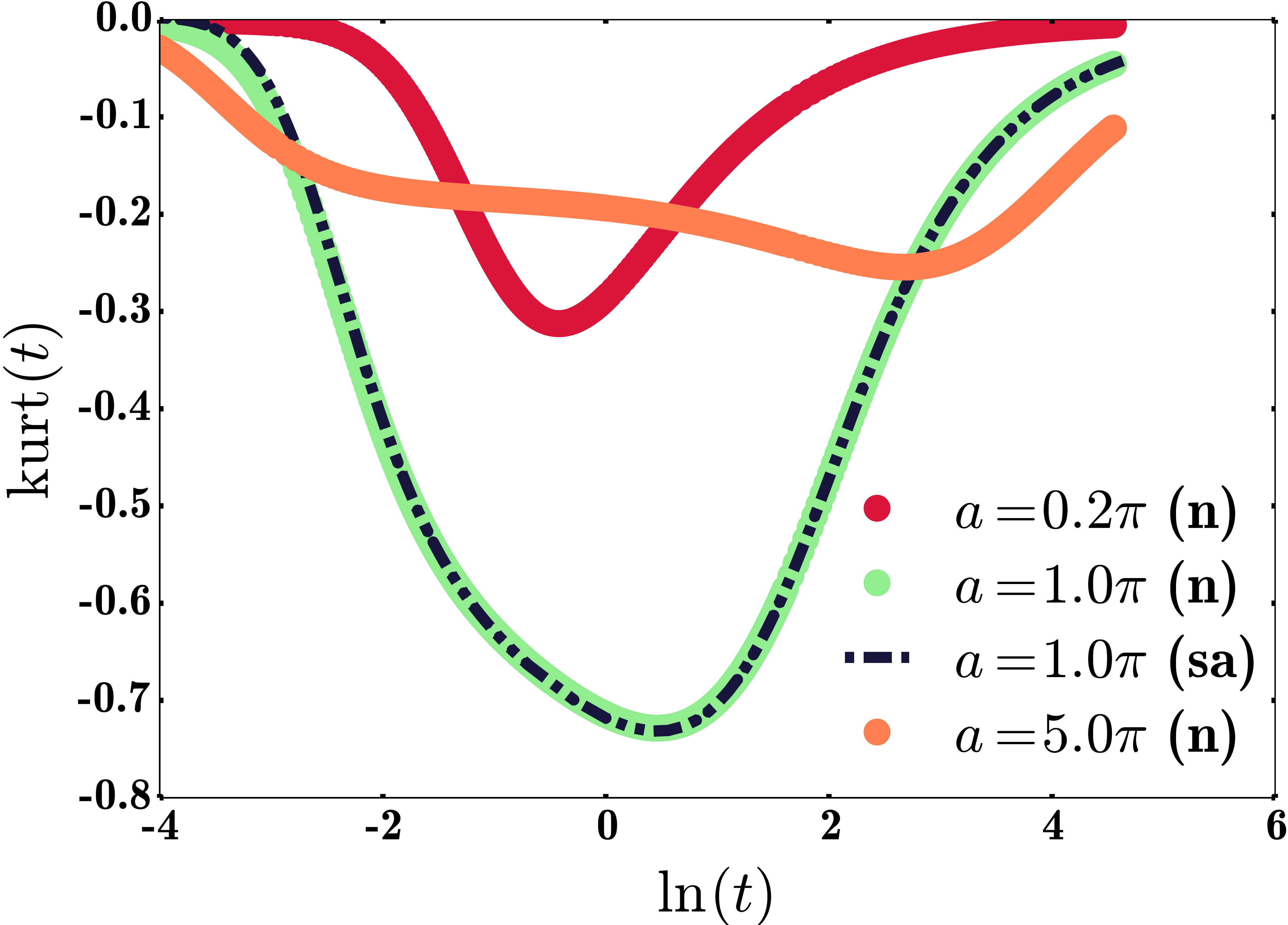}}
\caption{\rev{(Color online)} Kurtosis $\mathrm{kurt}(t)$ as a function of time $t$ for vanishing drift force $M=0$ and different maximum speeds $a$. The results are obtained by numerical forward-iteration of Eqs.~(\ref{eq_fp}) and (\ref{eq_init}), which is indicated by ``(n)''. For the case of $a=\pi$, results are also determined from semi-analytical calculations ``(sa)'' in Sec.~\ref{sec_analytical} via Eqs.~(\ref{eq_c2mp1})--(\ref{eq_e04}). The curves obtained in the two different ways show good agreement. Obviously, after an interval of significant non-Gaussian appearance, the magnitude of the kurtosis decays back to zero, which is the Gaussian value.}
\label{fig_kurtosis}
\end{figure}

\rev{In the case of linear (viscous) friction, the velocity and spatial distribution functions are of Gaussian shape \cite{risken1996fokker,zwanzig2001nonequilibrium}, see also Appendix~\ref{appendix_gauss}.} However, the nonlinear friction term Eq.~(\ref{eq_friction}) should lead to non-Gaussian spatial distribution functions on intermediate time scales, in analogy to what was previously observed for Coulomb friction \cite{menzel2011effect}. To detect deviations from Gaussian shape, it is useful to determine the kurtosis, which is related to the fourth moment of the distribution function, 
\begin{equation}\label{eq_kurt}
\mathrm{kurt} = \frac{\left\langle \big(x-\langle x\rangle\big)^4\right\rangle}{\left\langle\big( x-\langle x\rangle\big)^2\right\rangle^2}-3.
\end{equation}
In the Gaussian case, this expression for the kurtosis vanishes. 
Fig.~\ref{fig_kurtosis} depicts the time evolution of the kurtosis for different values of the maximum speed $a$ in our case of nonlinear friction. 
At early times, the kurtosis is small in magnitude, i.e.\ $c_0(x,t)$ is of nearly Gaussian shape. The reason is our initial condition Eq.~(\ref{eq_init}), where we start from a Gaussian distribution. Then, an obvious regime of non-Gaussian shape follows at intermediate times as a consequence of our nonlinear friction force. Finally, at later times, the central-limit theorem takes over \cite{kubo1991statistical} and the curves become Gaussian again. 

We observe that the kurtosis becomes negative at intermediate times. This indicates that the spatial distribution functions remain more compact and less extended than in the Gaussian case. In particular, it signals that long tails are suppressed. Such properties of the distribution function are naturally caused by the bounds on the possible range of velocities. High speeds are suppressed. Thus, long tails in the spatial distribution functions that result from high velocities are impeded when compared to the Gaussian case, where no bounds on the velocity range are present. 

Interestingly, the non-Gaussian behavior is most explicit in Fig.~\ref{fig_kurtosis} for the intermediate value of the maximum speed $a=1.0\pi$. We can readily identify a reason why for higher values of the maximum speed (here $a=5.0\pi$) the behavior becomes more Gaussian again. Around $v\approx0$, the argument of the tangent in Eq.~(\ref{eq_friction}) is smaller the higher is $a$. Then the tangent can be approximated by the first (linear) term of its expansion, see Eq.~(\ref{eq_linearized_friction}). This corresponds to a linear (viscous) friction term in agreement with Gaussian properties. Larger magnitudes of the velocity are necessary to enter the nonlinear friction regime, which are less likely under the same strength of the stochastic force $\gamma(t)$ in Eq.~(\ref{langevin_v}). We can also directly infer this behavior from Fig.~\ref{fig_fst_Meq0}, where the stationary velocity distribution $f_{\mathrm{st}}(v)$ for $a=5.0\pi$ is of nearly Gaussian shape in agreement with a nearly linear friction. 
\rev{Further comparison between the present results under nonlinear friction and the Gaussian results under linear (viscous) friction can be found in Appendix~\ref{appendix_gauss}.} 
Apart from that, we note that the buckled shape of the curve for $a=5.0\pi$ in Fig.~\ref{fig_kurtosis} could indicate different time scales for different processes determining the non-Gaussian behavior. At present, however, we cannot identify these different processes. 

In contrast to that, for a low magnitude of $a$ (see $a=0.2\pi$ in Fig.~\ref{fig_fst_Meq0}), there must be a different reason for the less explicit non-Gaussian behavior. We presume that here another mechanism is at work. Due to the high friction, the possibility to diffuse large distances is significantly hindered. Velocity magnitudes are restricted to lower values. The probability to find a particle remains much more concentrated around $x\approx0$. There, this leads to significantly higher statistical overlay in probabilities to encounter a particle. In turn, this adds to the central limit theorem, which favors the Gaussian character.

\subsection{Influence of a constant drift force}

\begin{figure}
\centerline{\includegraphics[width=7.5cm]{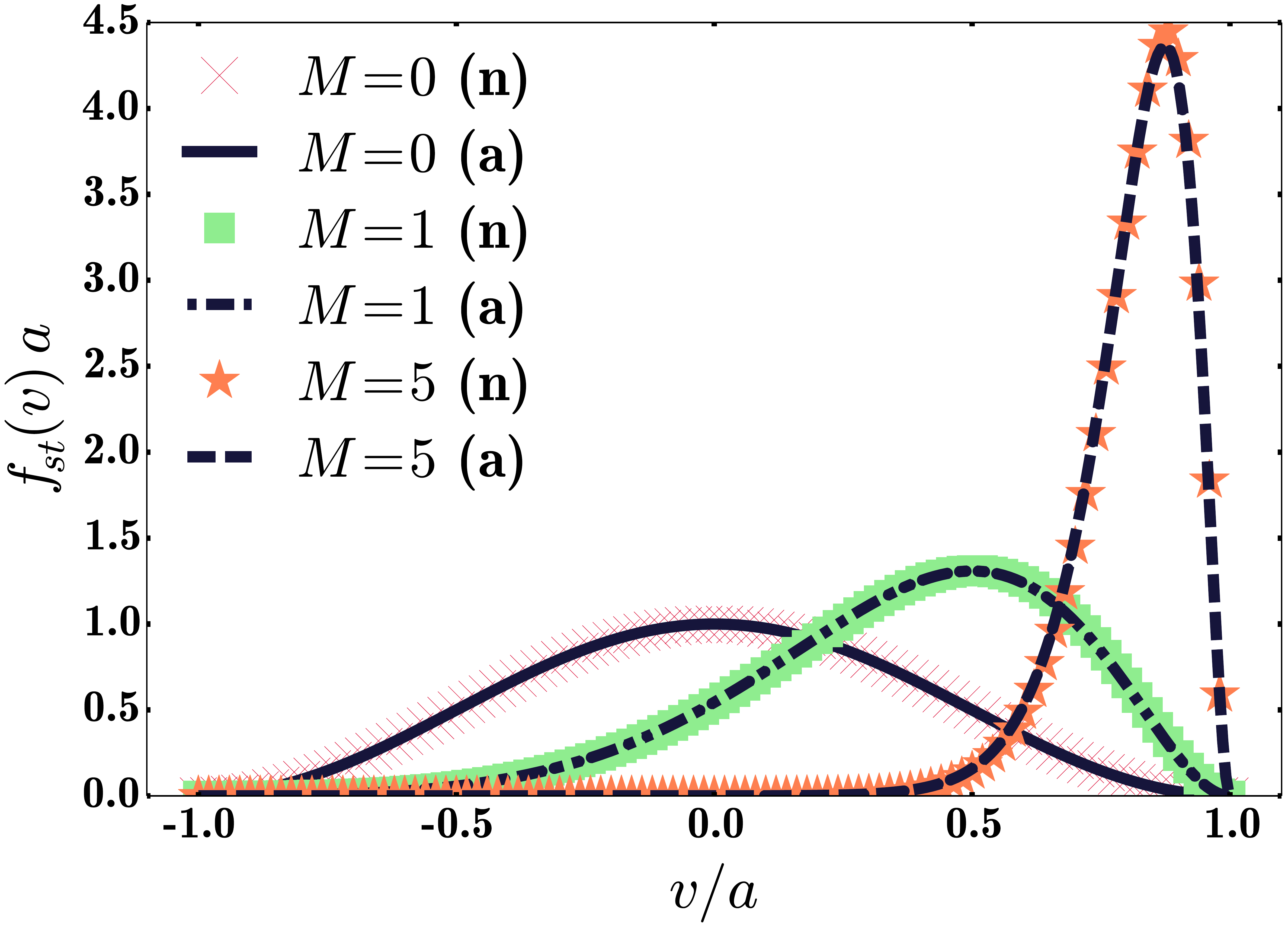}}
\caption{\rev{(Color online)} Comparison of $f_{\mathrm{st}}(v)\equiv f(v,t)$ as obtained via numerical solution of Eqs.~(\ref{eq_fp}) and (\ref{eq_init}) on the one hand, with analytical results from Eq.~(\ref{eq_fst}) on the other hand. Again, numerical results are marked by ``(n)'', analytical ones by ``(a)''. Here, different magnitudes of the drift force $M$ are considered for the same magnitude of the maximum speed $a=\pi$. As before, the time to determine $f_{\mathrm{st}}(v)\equiv f(v,t)$ was chosen as $t=5$, but is irrelevant. Numerical and analytical results show good agreement.}
\label{fig_fst_Mneq0}
\end{figure}

\begin{figure}
\centerline{\includegraphics[width=7.5cm]{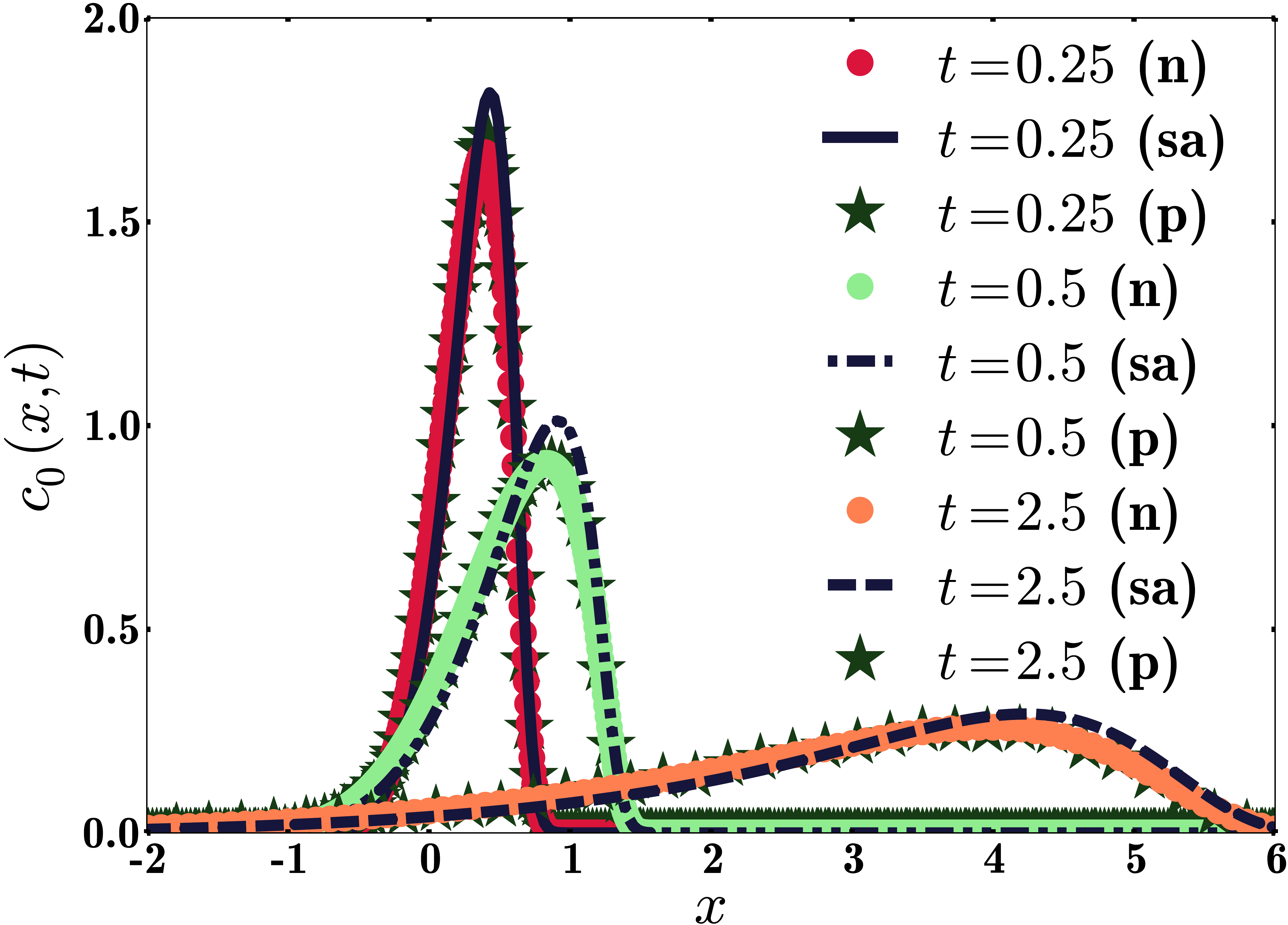}}
\caption{\rev{(Color online)} Time evolution of the spatial distribution function $c_0(x,t)$ for a non-vanishing drift force $M=1$ and a maximum speed $a=\pi$. The initial distribution is again given by Eq.~(\ref{eq_init}). Results obtained from direct numerical integration of the Fokker-Planck equation Eq.~(\ref{eq_fp}), via the semi-analytical approach combined with perturbation theory presented in Sec.~\ref{sec_analytical} in Eqs.~(\ref{eq_c2mp1})--(\ref{eq_c2m}), where Eqs.~(\ref{eq_e1})--(\ref{eq_e4}) are used as the right-hand sides of Eqs.~(\ref{eq_e01})--(\ref{eq_e04}), and from particle-based simulations, marked by ``(n)'', ``(sa)'', and ``(p)'', respectively, show good agreement. [Results from the particle-based simulations are all marked by the same symbol but are readily identified by the ``(n)''- and ``(sa)''-curves behind which they hide.] 
We remark that the drift force $M>0$ is directed to the right and use a nonsymmetric interval on the abscissa.}
\label{fig_timeseries_Mneq0}
\end{figure}

Next, we investigate the role of a nonvanishing drift force $M\neq0$. Fig.~\ref{fig_fst_Mneq0} again demonstrates that the numerical results obtained from Eqs.~(\ref{eq_fp}) and (\ref{eq_init}) for $f(v,t)\equiv f_{\mathrm{st}}(v)$ agree with the analytically calculated results from Eq.~(\ref{eq_fst}). 
From Fig.~\ref{fig_timeseries_Mneq0}, we can identify by eye the effect that the drift force has on the time evolution of the spatial distribution function $c_0(x,t)$. 
Naturally, over time, the distribution broadens and flattens as before for $M=0$. Now, however, due to the imposed drift $M\neq0$, the center of the distribution shifts along the direction of $M$. Moreover, the distribution becomes asymmetric. Again, particle-based simulations confirm our numerical results. In the following, we quantify our observations by the corresponding statistical measures. 

\begin{figure}
\centerline{\includegraphics[width=7.5cm]{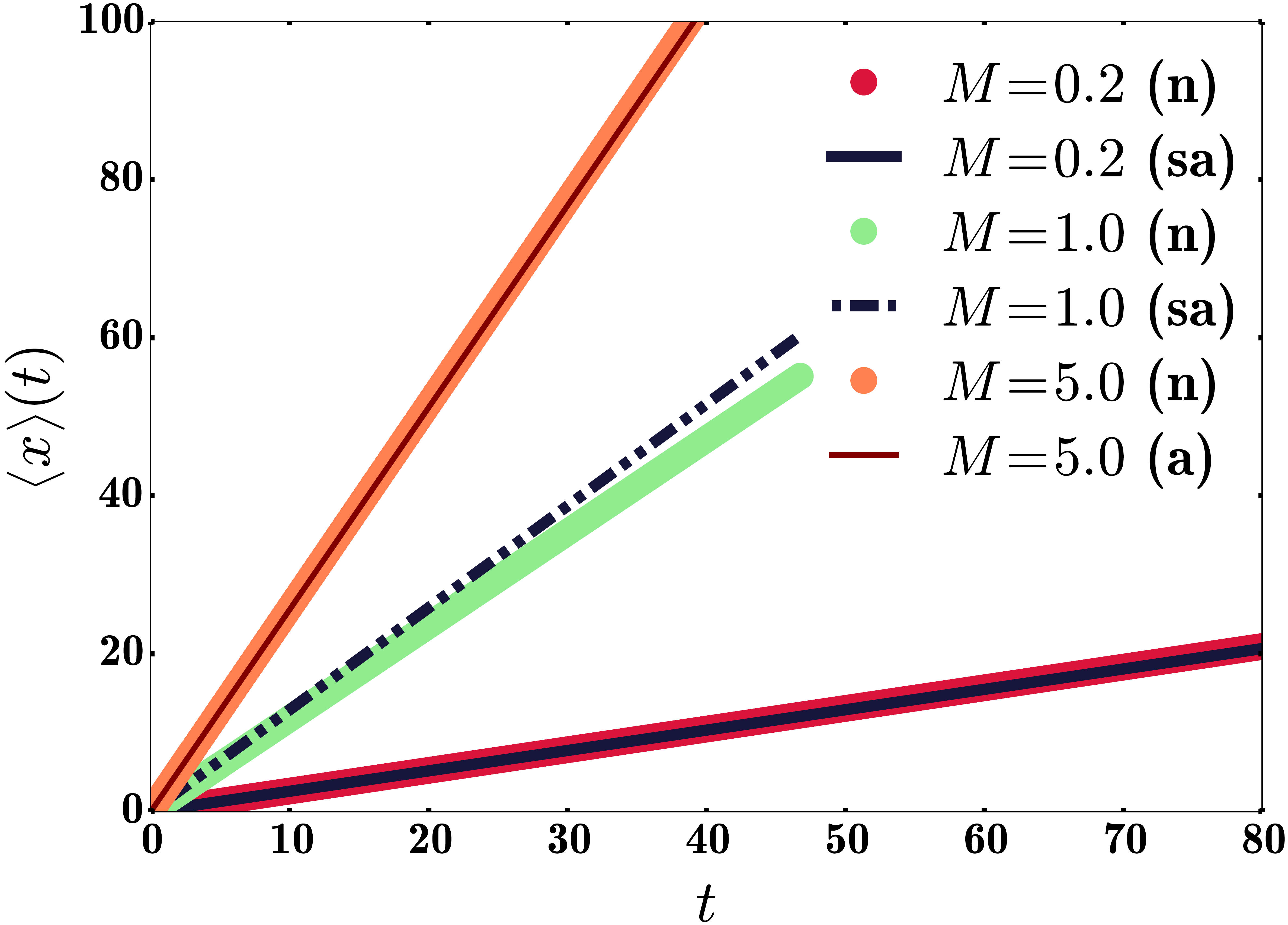}}
\caption{\rev{(Color online)} Drifting average position $\langle x\rangle(t)$ as a function of time $t$ for a maximum speed $a=\pi$ and for different drift forces $M$. We depict results obtained via numerical forward-iteration of Eqs.~(\ref{eq_fp}) and (\ref{eq_init}) marked by ``(n)'' as well as semi-analytical curves ``(sa)'' determined via our perturbational analysis in Sec.~\ref{sec_analytical} by forward-iteration of Eqs.~(\ref{eq_c2mp1})--(\ref{eq_e04}) combined with Eqs.~(\ref{eq_e1})--(\ref{eq_e4}). 
For $M=5.0$, where the perturbational analysis breaks down, we include instead the analytical result derived in Eq.~(\ref{eq_drift_analytical}) and label it by ``(a)''. The curves obtained in the different ways show good agreement.}
\label{fig_drift_Mneq0}
\end{figure}

Under the imposed $M$, the average particle position $\langle x\rangle(t)$ starts to drift into the direction of the applied force. We can directly obtain $\langle x\rangle(t)$ from our numerical calculations via 
\begin{equation}\label{eq_drift_numerical}
\langle x\rangle(t) = \int_{-\infty}^{\infty}\mathrm{d}x \int_{-a}^{a}\mathrm{d}v \:x\,f(x,v,t). 
\end{equation}
Corresponding numerical results for different values of $M$ are displayed in Fig.~\ref{fig_drift_Mneq0} and show the expected linear increase of $\langle x\rangle(t)$ with time. 

In addition to that, for $a=\pi$, we can calculate $\langle x\rangle(t)$ analytically. As noted before, integrating Eq.~(\ref{eq_fp}) and our initial condition Eq.~(\ref{eq_init}) over $x$ leads us to Eq.~(\ref{eq_fvt}). Next, multiplying Eq.~(\ref{eq_fp}) by $x$ and integrating over $x$ and $v$, together with Eq.~(\ref{eq_fst}) we find 
\begin{equation}
\partial_t\left[\langle x\rangle(t)\right] = \int_{-a}^a\mathrm{d}v\:v\,f_{\mathrm{st}}(v) = \partial_M\ln Z. 
\end{equation}
For $a=\pi$, the integral can be solved analytically, and we obtain 
\begin{equation}
\langle x\rangle (t) = \frac{\pi\left(M+M^3\right)\coth\left(\pi M\right)-3M^2-1}{M+M^3}\,t. 
\label{eq_drift_analytical}
\end{equation}
We find the intuitive limits $\lim_{M\rightarrow0}\langle x\rangle (t)=0$ and $\lim_{M\rightarrow\infty}\langle x\rangle (t)=\pi t$. Example curves are indicated in Fig.~\ref{fig_drift_Mneq0}. 

\begin{figure}
\centerline{\includegraphics[width=7.5cm]{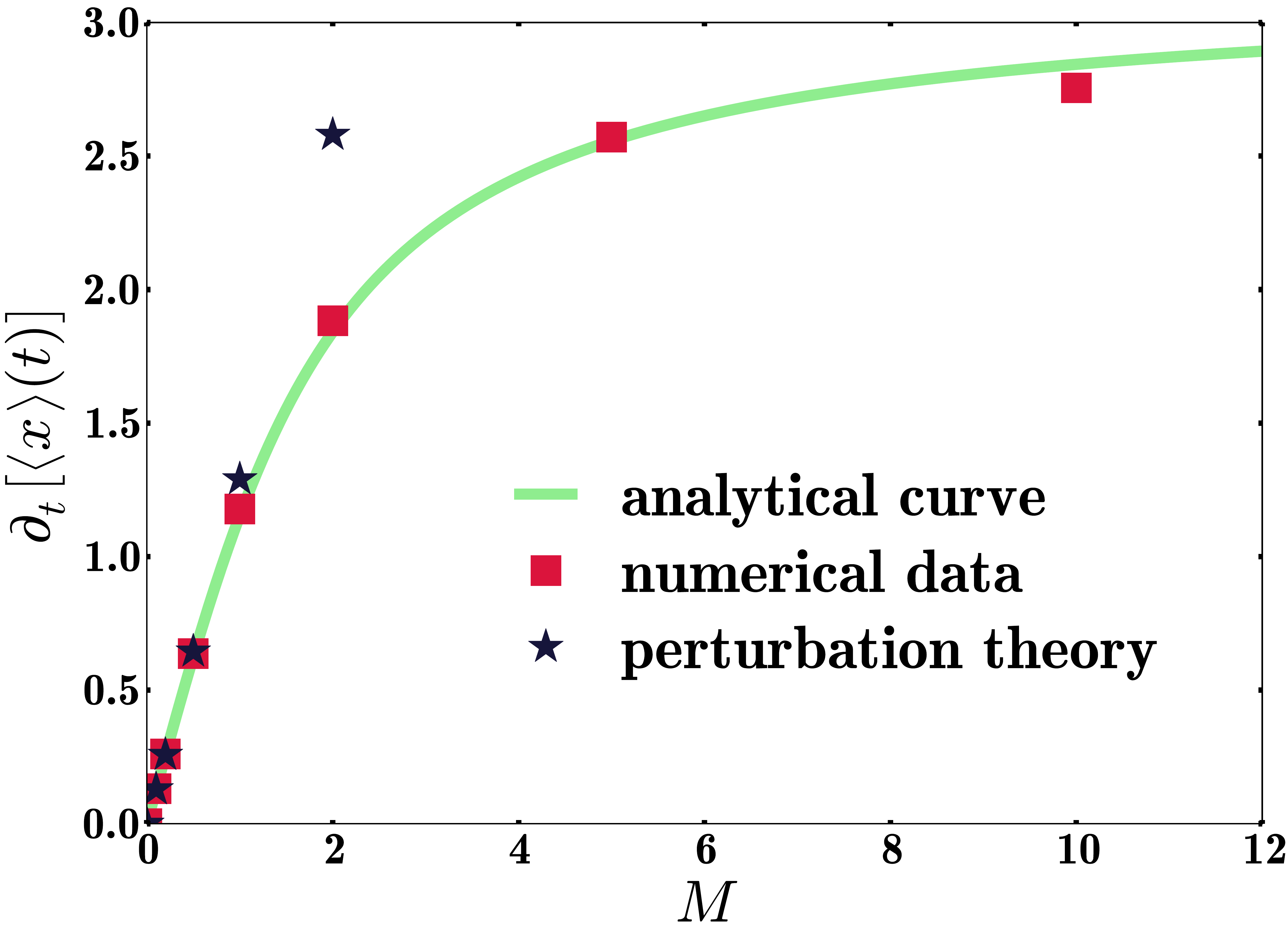}}
\caption{\rev{(Color online)} Comparison between three different ways of determining the drift velocity $\partial_t\left[\left\langle x\right\rangle(t)\right]$ for $a=\pi$ as a function of the drift force $M$: analytically via Eq.~(\ref{eq_drift_analytical}); numerically by forward-iteration of Eq.~(\ref{eq_fp}) and via Eq.~(\ref{eq_drift_numerical}); and from our perturbational analysis in Sec.~\ref{sec_analytical} by forward-iteration of Eqs.~(\ref{eq_c2mp1})--(\ref{eq_e04}) combined with Eqs.~(\ref{eq_e1})--(\ref{eq_e4}).} 
\label{fig_driftvelocity}
\end{figure}

Fig.~\ref{fig_driftvelocity} contains the resulting drift velocities $\partial_t\left[\left\langle x\right\rangle(t)\right]$ for $a=\pi$. 
The analytical results follow directly from Eq.~(\ref{eq_drift_analytical}) as $\langle x \rangle(t)/t$. 
We find that our numerical data obtained via Eq.~(\ref{eq_drift_numerical}) well reproduce the analytical curve. The small deviations for large drift forces ($M=10$) are attributed to our finite-difference discretization scheme. 

\begin{figure}
\centerline{\includegraphics[width=7.5cm]{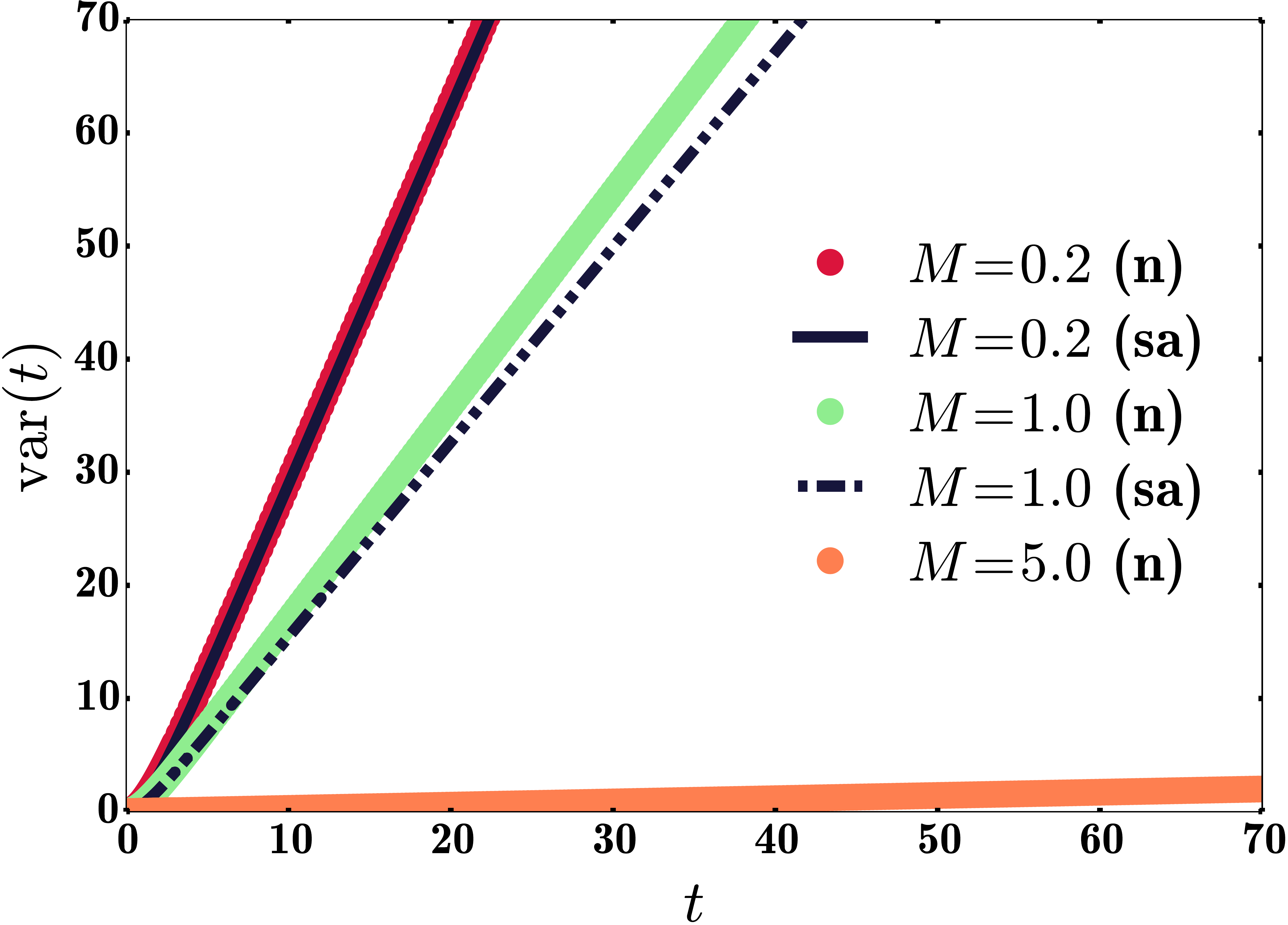}}
\caption{\rev{(Color online)} Variance $\mathrm{var}(t)$ as a function of time $t$ for a maximum speed $a=\pi$ and for different drift forces $M$. Depicted results originate, on the one hand, from numerical forward-iteration of Eqs.~(\ref{eq_fp}) and (\ref{eq_init}) marked by ``(n)''; on the other hand, they are calculated semi-analytically ``(sa)'' via our perturbational analysis in Sec.~\ref{sec_analytical} by Eqs.~(\ref{eq_c2mp1})--(\ref{eq_e04}) combined with Eqs.~(\ref{eq_e1})--(\ref{eq_e4}). 
The curves obtained in the different ways show good agreement. Increasing $M>0$ hinders the broadening of the spatial distributions.}
\label{fig_var_Mneq0}
\end{figure}

Again, after an initial transient has decayed, the variance of the spatial distributions as defined in Eq.~(\ref{eq_var}) grows linearly in time, see Fig.~\ref{fig_var_Mneq0}. 
The increase in the variance, i.e.\ the broadening of the distributions, is significantly hindered for increasing magnitude of $M$. This observation agrees with the curves shown in Fig.~\ref{fig_fst_Mneq0} for the stationary velocity distributions. At higher $M$, the velocity distributions become more narrow because the drift force $M>0$ pushes them towards the positive edge of maximum velocity. This narrows the spectrum of different available velocities that would broaden the spatial distribution. 

\begin{figure}
\centerline{\includegraphics[width=7.5cm]{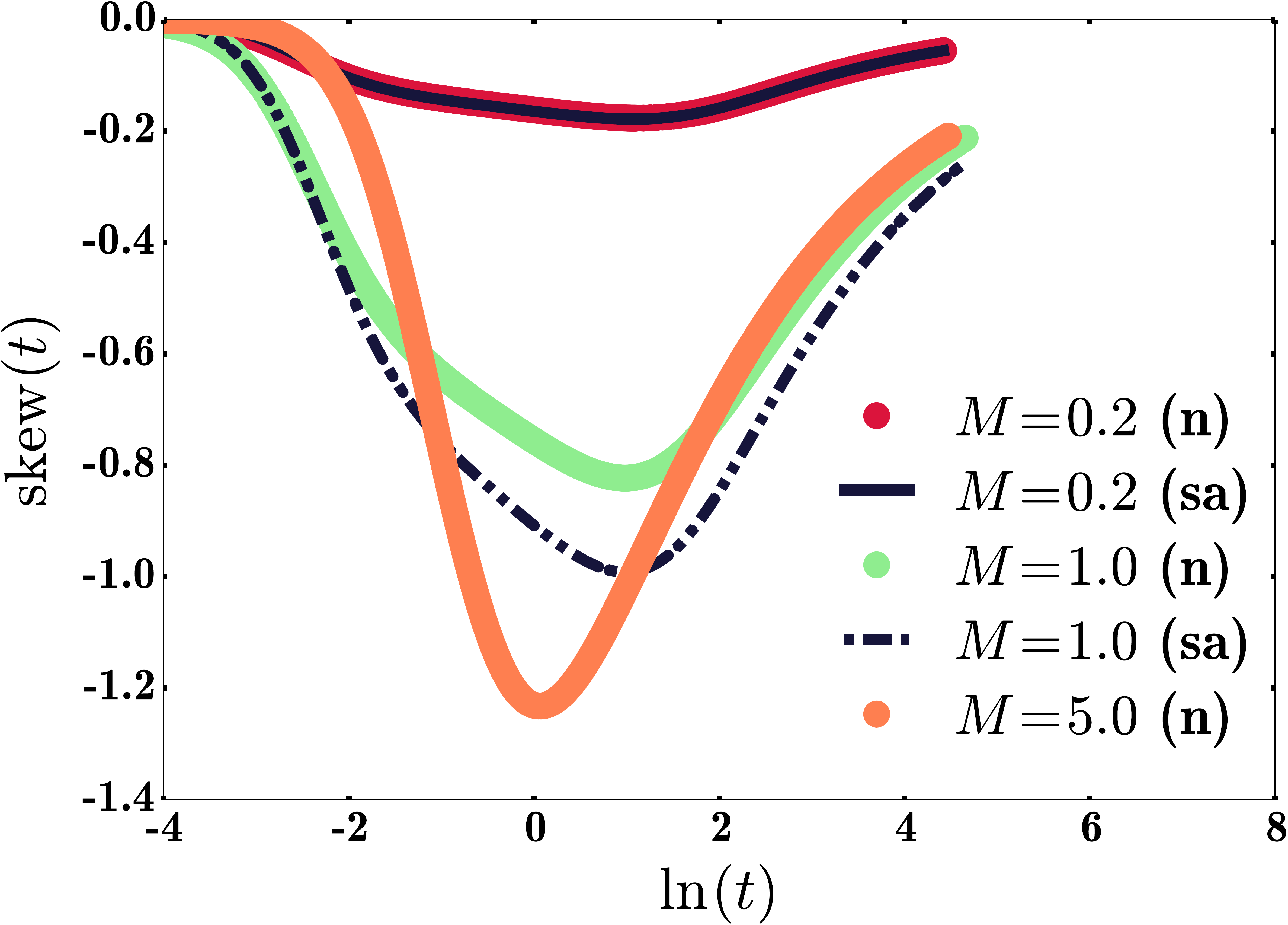}}
\caption{\rev{(Color online)} Skewness $\mathrm{skew}(t)$ as a function of time $t$ for a maximum speed $a=\pi$ and for different drift forces $M$. We depict results, on the one hand, from numerical forward-iteration of Eqs.~(\ref{eq_fp}) and (\ref{eq_init}), marked by ``(n)'', and, on the other hand, from semi-analytical calculations ``(sa)'' via our perturbational analysis in Sec.~\ref{sec_analytical} by Eqs.~(\ref{eq_c2mp1})--(\ref{eq_e04}) combined with Eqs.~(\ref{eq_e1})--(\ref{eq_e4}). 
As before, the curves obtained in the different ways show good agreement for not too high magnitudes of $M$. Obviously, after an explicit interval of asymmetry in the spatial distribution curves, the skewness decays again at larger times.}
\label{fig_skew_Mneq0}
\end{figure}

The asymmetry arising in the curves of $c_0(x,t)$ in Fig.~\ref{fig_timeseries_Mneq0} can be quantified via the skewness, 
\begin{equation}
\mathrm{skew} = \frac{\left\langle \big(x-\langle x\rangle\big)^3\right\rangle}{\left\langle\big( x-\langle x\rangle\big)^2\right\rangle^{\frac{3}{2}}}. 
\end{equation}
We show examples for the time evolution of the skewness under imposed drift in Fig.~\ref{fig_skew_Mneq0}. 
These results identify a pronounced interval of asymmetry in the spatial distribution functions. The skewness is negative in all cases, which implies that the curves lean to the right. This agrees with a positive $M>0$ that drives the distribution towards the positive $x$ direction, see Fig.~\ref{fig_timeseries_Mneq0}. At longer times, the magnitude of the skewness decays again, i.e.\ the curves return to more symmetric shapes. 

\begin{figure}
\centerline{\includegraphics[width=7.5cm]{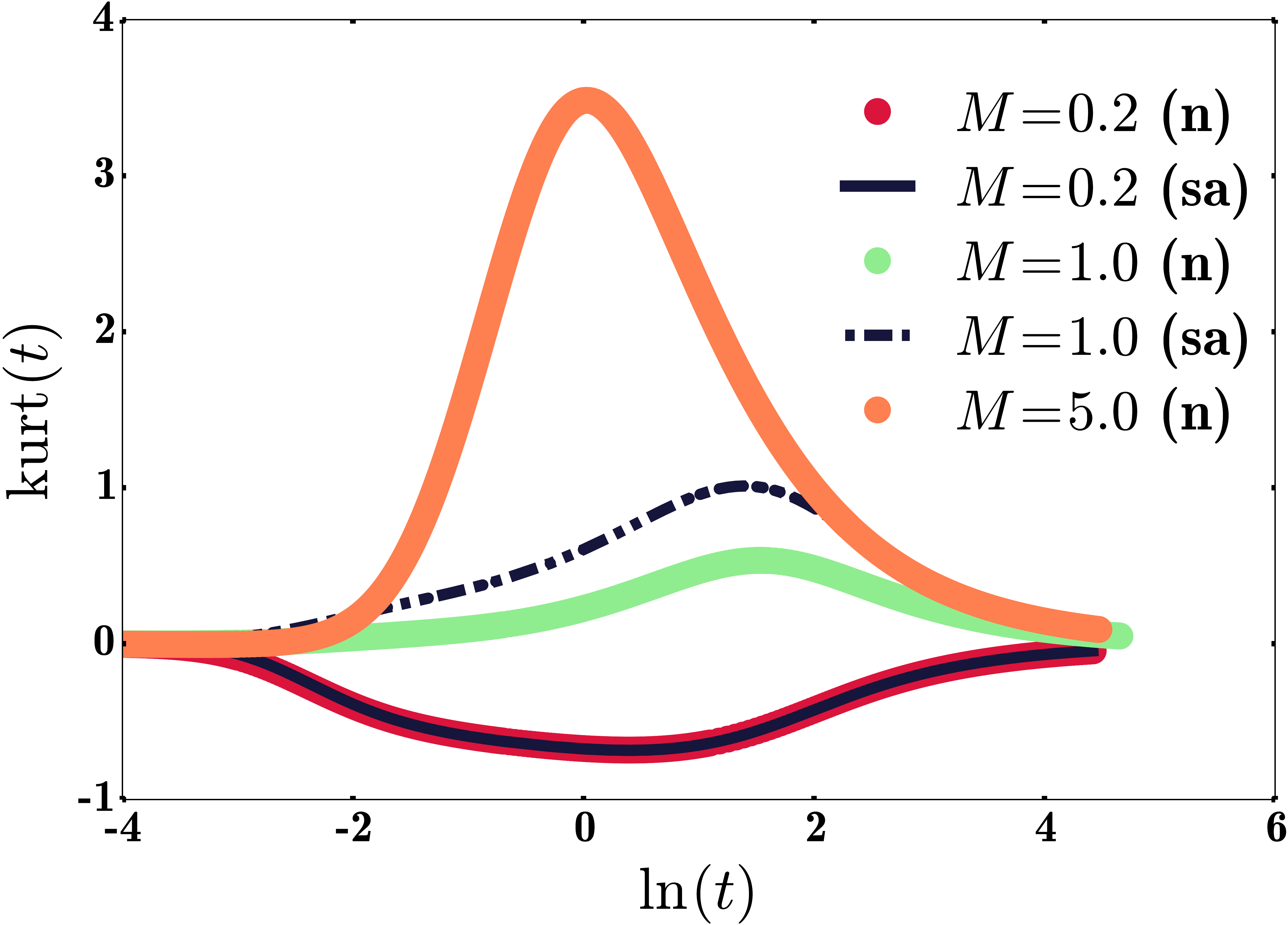}}
\caption{\rev{(Color online)} Kurtosis $\mathrm{kurt}(t)$ as a function of time $t$ for a maximum speed $a=\pi$ and for different drift forces $M$. Again, results obtained from numerical forward-iteration of Eqs.~(\ref{eq_fp}) and (\ref{eq_init}), marked by ``(n)'', are depicted on the one hand. On the other hand, results are determined from semi-analytical calculations ``(sa)'' via our perturbational analysis in Sec.~\ref{sec_analytical} by Eqs.~(\ref{eq_c2mp1})--(\ref{eq_e04}) combined with Eqs.~(\ref{eq_e1})--(\ref{eq_e4}). 
As before, the curves obtained in the two different ways show good agreement for not too high magnitudes of $M$. Obviously, after an interval of pronounced non-Gaussian shape, the magnitude of the kurtosis decays back to zero, i.e.\ the Gaussian value.}
\label{fig_kurt_Mneq0}
\end{figure}

Finally, we measure the kurtosis defined in Eq.~(\ref{eq_kurt}) under drift $M\neq0$ and depict the results in Fig.~\ref{fig_kurt_Mneq0}. 
During an intermediate time interval, the kurtosis in Fig.~\ref{fig_kurt_Mneq0} significantly deviates from zero, which indicates non-Gaussian shape. For lower magnitude of $M$, the kurtosis is still negative as in the case of $M=0$. That is, the hindered motion due to the nonlinear increase in friction at higher speeds still leads to a more concentrated particle distribution when compared to corresponding cases of linear friction. Interestingly, however, higher values of the drift force $M\neq0$ counteract this trend. The kurtosis becomes positive, which implies more extended shapes (longer tails) than for comparable Gaussian distributions. At longer times, the magnitude of the kurtosis decays to zero again, which is the Gaussian value.

\section{Analytical treatment}\label{sec_analytical}

As an appealing aspect of our model, significant analytical progress is possible for a specific value of the rescaled maximum speed $a$ in Eq.~(\ref{eq_fp}): 
\begin{equation}
a={\pi}. 
\end{equation}
We keep this specific choice during the whole section. 

Eq.~(\ref{eq_fp}) can be written in the symbolic form $\partial_t f=Lf$, with L the Fokker-Planck operator. To make analytical progress, we perform the conventional transformation $\bar{L}=\sqrt{f_{\mathrm{st}}}^{\,-1}L\sqrt{f_{\mathrm{st}}}$ and $\bar{f}=\sqrt{f_{\mathrm{st}}}^{\,-1}f$ \cite{risken1996fokker} by dividing Eq.~(\ref{eq_fp}) through $\sqrt{f_{\mathrm{st}}}$, where $f_{\mathrm{st}}$ is given by Eq.~(\ref{eq_fst}). Using $a=\pi$ allows to simplify the resulting equation to 
\begin{equation}\label{eq_fbar}
\partial_t \bar{f} = -v\partial_x\bar{f} + \partial_v^2\bar{f} + \frac{1}{4}\left(1-M^2\right)\bar{f} + \frac{1}{2}M\tan\left(\frac{v}{2}\right)\bar{f}. 
\end{equation}

\pagebreak

\subsection{Bare diffusion under nonlinear friction}

In the absence of an externally imposed drift force, i.e.\ $M=0$, Eq.~(\ref{eq_fbar}) simplifies to
\begin{equation}\label{eq_fp0_barf}
\partial_t \bar{f} = -v\partial_x\bar{f} + \partial_v^2\bar{f} + \frac{1}{4}\bar{f}. 
\end{equation}
We read off the reversible part $\bar{L}_{\mathrm{rev}}=-v\,\partial_x$ and the irreversible part $\bar{L}_{\mathrm{ir}}=\partial_v^2+\frac{1}{4}$ of the remaining Fokker-Planck operator. In general, the advantage of this procedure \cite{risken1996fokker} is that $\bar{L}_{\mathrm{ir}}$ is now Hermitian and its eigenvalues and eigenfunctions can be determined. We denote this problem in the form 
\begin{equation}\label{eq_Lir}
\bar{L}_{\mathrm{ir}}\psi_{\mu}(v) = -\mu\,\psi_{\mu}(v). 
\end{equation}
This formulation reveals the well-known general connection between the irreversible part of the Fokker-Planck equation and a corresponding Schr\"odinger equation \cite{risken1996fokker}. Here, we obtain
\begin{equation}
\left({}-\partial_v^2-\frac{1}{4}\right)\psi_{\mu}(v)=\mu\,\psi_{\mu}(v). 
\end{equation}
We recall that our velocities are restricted to the interval $(-a,a)$ due to the divergence of the tangent in the frictional force Eq.~(\ref{eq_friction}). Outside this interval, our velocity distribution vanishes. Thus, in our case, we find a connection to the Schr\"odinger equation for a particle in an infinite square-well potential that is constant within $(-\pi,\pi)$ and infinite outside \cite{risken1996fokker}. The orthonormal eigenfunctions and eigenvalues corresponding to this situation are well known, and we label them according to their frequencies, respectively: 
\begin{equation}\label{eq_0_2np1}
\psi_{2n+1}^{(0)}(v) = \frac{1}{\sqrt{\pi}}\,\cos\left(\frac{2n+1}{2}v\right), 
\quad \mu_{2n+1}=n(n+1), 
\end{equation}
where $n=0,1,2,\dots$, and
\begin{equation}\label{eq_0_2n}
\psi_{2n}^{(0)}(v) = \frac{1}{\sqrt{\pi}}\,\sin\left(\frac{2n}{2}v\right), 
\quad \mu_{2n}=n^2-\frac{1}{4}, 
\end{equation}
with $n=1,2,\dots$. 

As a benefit, we can now separate the spatial and velocity dependences in $\bar{f}(x,v,t)$ by expanding $\bar{f}$ in the eigenfunctions of $\bar{L}_{\mathrm{ir}}$: 
\begin{equation}\label{eq_barf}
\bar{f}(x,v,t) = \sum_{n=0}^{\infty}c_{2n+1}(x,t)\psi^{(0)}_{2n+1}(v) + \sum_{n=1}^{\infty}c_{2n}(x,t)\psi^{(0)}_{2n}(v).
\end{equation}
Realizing that $\sqrt{f_{\mathrm{st}}(v)}=\psi_{2\cdot0+1}^{(0)}(v)$, we notice that our searched-for spatial distribution function $\int_{-\pi}^{\pi}\mathrm{d}v\,f(x,v,t)=c_{2\cdot0+1}(x,t)$. 

Next, we include the reversible operator $\bar{L}_{\mathrm{rev}}=-v\,\partial_x$. This leads to $v$-dependent terms of the kind $v\,\psi_{2n+1}^{(0)}$ and $v\,\psi_{2n}^{(0)}$. To stay within our framework of orthonormal eigenfunctions, we expand these contributions in terms of $\psi_{2n+1}^{(0)}$ and $\psi_{2n}^{(0)}$: 
\begin{eqnarray}
v\,\psi_{2n+1}^{(0)} &=& 
  \sum_{k=0}^{\infty}\left\langle\psi_{2k+1}^{(0)}\Big|\,v\,\psi_{2n+1}^{(0)}\right\rangle\psi_{2k+1}^{(0)} \nonumber\\
  &&{}\quad + \sum_{k=1}^{\infty}\left\langle\psi_{2k}^{(0)}\Big|\,v\,\psi_{2n+1}^{(0)}\right\rangle\psi_{2k}^{(0)}, 
\label{eq_expansion_v2np1}\\
v\,\psi_{2n}^{(0)} &=& \sum_{k=0}^{\infty}\left\langle\psi_{2k+1}^{(0)}\Big|\,v\,\psi_{2n}^{(0)}\right\rangle\psi_{2k+1}^{(0)} \nonumber\\
  &&{}\quad + \sum_{k=1}^{\infty}\left\langle\psi_{2k}^{(0)}\Big|\,v\,\psi_{2n}^{(0)}\right\rangle\psi_{2k}^{(0)}. 
\label{eq_expansion_v2n}
\end{eqnarray}
The scalar product $\langle\cdot|\cdot\rangle$ in our case is simply given by the integral $\int_{-\pi}^{\pi}\cdot\cdot\, \mathrm{d}v$. 
Due to the symmetry properties of the eigenfunctions, we immediately obtain 
\begin{equation}
\left\langle\psi_{2k+1}^{(0)}\Big|\,v\,\psi_{2n+1}^{(0)}\right\rangle 
= 0 =
\left\langle\psi_{2k}^{(0)}\Big|\,v\,\psi_{2n}^{(0)}\right\rangle. 
\end{equation}
The only nonvanishing integrals can be solved analytically and read
\begin{eqnarray}
J_{nk} &:=& \left\langle\psi_{2k}^{(0)}\Big|\,v\,\psi_{2n+1}^{(0)}\right\rangle
=\left\langle\psi_{2n+1}^{(0)}\Big|\,v\,\psi_{2k}^{(0)}\right\rangle 
\nonumber\\
&=& \frac{1}{\pi}\int_{-\pi}^{\pi}\mathrm{d}v\,\cos\left(\frac{2n+1}{2}v\right)\sin\left(kv\right)\,v
\nonumber\\
&=&{}-\frac{32\,(-1)^{n+k}\,k\,(1+2n)}{\pi\left[(1+2n)^2-4k^2\right]^2}. 
\end{eqnarray}

Now inserting Eqs.~(\ref{eq_Lir}) as well as (\ref{eq_0_2np1})--(\ref{eq_expansion_v2n}) 
back into Eq.~(\ref{eq_fp0_barf}) and projecting onto the eigenfunctions $\psi_{2m+1}^{(0)}(v)$ and $\psi_{2m}^{(0)}(v)$, we obtain a coupled dynamic set of equations for the expansion coefficients: 
\begin{widetext}
\begin{eqnarray}
\partial_t\, c_{2m+1}(x,t) &=& {}-\sum_{n=0}^{\infty}e_{2m+1,2n+1}\,\partial_x\,c_{2n+1}(x,t) -\sum_{n=1}^{\infty}e_{2m+1,2n}\,\partial_x\,c_{2n}(x,t) 
-\mu_{2m+1}\,c_{2m+1}(x,t), 
\label{eq_c2mp1}
\end{eqnarray}
\pagebreak
\begin{eqnarray}
\partial_t\, c_{2m}(x,t) &=& {}-\sum_{n=0}^{\infty}e_{2m,2n+1}\,\partial_x\,c_{2n+1}(x,t) -\sum_{n=1}^{\infty}e_{2m,2n}\,\partial_x\,c_{2n}(x,t) 
-\mu_{2m}\,c_{2m}(x,t), 
\label{eq_c2m}
\end{eqnarray}
\end{widetext}
where
\begin{eqnarray}
e_{2m+1,2n+1} &=& \left\langle\psi_{2m+1}^{(0)}\Big|\,v\,\psi_{2n+1}^{(0)}\right\rangle, 
\label{eq_e01}\\
 e_{2m+1,2n} &=& \left\langle\psi_{2m+1}^{(0)}\Big|\,v\,\psi_{2n}^{(0)}\right\rangle, \\
e_{2m,2n+1} &=& \left\langle\psi_{2m}^{(0)}\Big|\,v\,\psi_{2n+1}^{(0)}\right\rangle, \\
e_{2m,2n} &=& \left\langle\psi_{2m}^{(0)}\Big|\,v\,\psi_{2n}^{(0)}\right\rangle. 
\label{eq_e04}
\end{eqnarray}
In our case $e_{2m+1,2n+1}=e_{2m,2n}=0$, while $e_{2m+1,2n}=e_{2n,2m+1}=J_{mn}$. Eqs.~(\ref{eq_c2mp1}) and (\ref{eq_c2m}) can be rewritten in the form of a matrix equation
\begin{equation}\label{eq_matrix}
\partial_t\,\mathbf{c}(x,t) = {}-\mathbf{e}\cdot\partial_x\,\mathbf{c}(x,t) -\bm{\mu}\cdot\mathbf{c}(x,t). 
\end{equation}
Here, $\mathbf{c}(x,t)$ is a vector composed of the expansion coefficients $c_{2m+1}(x,t)$ and $c_{2m}(x,t)$ in this order and for increasing $m$, respectively. $\mathbf{e}$ consists of the entries given by Eqs.~(\ref{eq_e01})--(\ref{eq_e04}), ordered along lines and columns in analogy to the vector $\mathbf{c}(x,t)$; it is real and symmetric. Finally, $\bm{\mu}$ is a diagonal matrix with the eigenvalues $\mu_{2m+1}$ and $\mu_{2m}$ listed along its diagonal. 

Our initial condition in Eq.~(\ref{eq_init}) sets $\mathbf{c}(x,t=0)$: we find that all $c_i(x,t=0)=0$, except for
\begin{equation}\label{eq_c0init}
c_0(x,t=0) = \frac{1}{\sqrt{\pi}\sigma}\,\exp\left\{-\left(\frac{x}{\sigma}\right)^2\right\}. 
\end{equation}
Eq.~(\ref{eq_matrix}) can then be solved by numerically iterating this initial state forward in time. 

In practice, the system of equations must be cut at a certain finite order. As a benefit of this whole procedure, only a relatively low number of equations is needed to obtain decent results. For example, the semi-analytical curves in Fig.~\ref{fig_timeseries_Meq0} were determined by including expansion coefficients $c_{2m+1}(x,t)$ and $c_{2m}(x,t)$ up to the order $m=5$ only in Eqs.~(\ref{eq_c2mp1}) and (\ref{eq_c2m}). 
There is good agreement with the results obtained by direct numerical forward-iteration of the Fokker-Planck equation, Eqs.~(\ref{eq_fp}) and (\ref{eq_init}). 
Likewise, calculating the first four moments leads to results consistent with the direct numerical solution. The first and the third moments vanish, the variance and kurtosis are compared to the fully numerical approach in Figs.~\ref{fig_var} and \ref{fig_kurtosis}, respectively. 
We checked that including more expansion coefficients does not noticeably modify the obtained results. 

The whole procedure leads to a significant numerical speed-up: our expansion coefficients are restricted to the one-dimensional $x$ direction. In contrast to that, our initial Eq.~(\ref{eq_fp}) needed to be iterated on the two-dimensional $x$-$v$ plane.

\subsection{Influence of a constant drift force}

Next, we additionally take into account the externally imposed drift force $M$, i.e.\ we consider the full Eq.~(\ref{eq_fbar}). It is straightforward to include the factor $(1-M^2)$ appearing in Eq.~(\ref{eq_fbar}). This factor enters the irreversible part of the Fokker-Planck operator, which becomes $\bar{L}_{\mathrm{ir}}=\partial_v^2+\frac{1}{4}(1-M^2)$. Thus, within the corresponding Schr\"odinger equation, its only effect is to shift the bottom level of the square-well potential from $-\frac{1}{4}$ to $-\frac{1}{4}(1-M^2)$. As a consequence, we obtain the same functional form of the eigenfunctions as in Eqs.~(\ref{eq_0_2np1}) and (\ref{eq_0_2n}). The only difference arises in the eigenvalues, which are shifted by $\frac{1}{4}M^2$: 
\begin{eqnarray}
\mu_{2n+1}^{(0)} = n(n+1)+\frac{1}{4}M^2, &&\quad n=0,1,2,\dots,
\label{eq_mu0_2np1}
\\
\mu_{2n}^{(0)} = n^2-\frac{1}{4}+\frac{1}{4}M^2, &&\quad n=1,2,\dots.
\label{eq_mu0_2n}
\end{eqnarray}

The second contribution due to the externally imposed drift force, i.e.\ the last term in Eq.~(\ref{eq_fbar}), is more difficult to handle. We include it within the framework of quantum-mechanical perturbation theory \cite{landau1077quantum}, which for consistency is conducted up to the second order in $M$. Thus we expand the corrections to the eigenvalues and eigenfunctions up to second order in $M$, which implies
\begin{eqnarray}
\mu_{2n+1} &=& \mu_{2n+1}^{(0)} + M\mu_{2n+1}^{(1)} + M^2\mu_{2n+1}^{(2)} +\dots, 
\\[.1cm]
\mu_{2n} &=& \mu_{2n}^{(0)} + M\mu_{2n}^{(1)} + M^2\mu_{2n}^{(2)} +\dots, 
\\[.1cm]
\psi_{2n+1} &=& \psi_{2n+1}^{(0)} + M\psi_{2n+1}^{(1)} + M^2\psi_{2n+1}^{(2)} +\dots, 
\label{eq_psi2np1}
\\[.1cm]
\psi_{2n} &=& \psi_{2n}^{(0)} + M\psi_{2n}^{(1)} + M^2\psi_{2n}^{(2)} +\dots. 
\label{eq_psi2n}
\end{eqnarray}
Following the quantum-mechanical analogy, our ``Hamiltonian'' is expanded as
\begin{equation}
H = H_0+MH_1, 
\end{equation}
where
\begin{eqnarray}
H_0 &=& -\bar{L}_{\mathrm{ir}} = {}-\partial_v^2-\frac{1}{4}\left(1-M^2\right), \\
H_1 &=& {}-\frac{1}{2\,}\tan\left(\frac{v}{2}\right). 
\label{eq_H1}
\end{eqnarray}
[Strictly speaking, $H_0$ also contains a contribution of order $M^2$. Yet, since its effect can be determined exactly and does not affect the eigenfunctions, we include this part into $H_0$ and its influence into $\mu_{2n+1}^{(0)}$ and $\mu_{2n}^{(0)}$, see Eqs.~(\ref{eq_mu0_2np1}) and (\ref{eq_mu0_2n}), respectively.] The lowest (zeroth) order corresponding to the unperturbed Hamiltonian $H_0$ has already been solved, with the eigenfunctions listed in Eqs.~(\ref{eq_0_2np1}) and (\ref{eq_0_2n}) as well as the eigenvalues given by Eqs.~(\ref{eq_mu0_2np1}) and (\ref{eq_mu0_2n}).

Next, the first-order corrections to the eigenvalues are given by 
\begin{eqnarray}
M\mu_{2n+1}^{(1)} & = & M\left\langle \psi_{2n+1}^{(0)}\Big|\,H_1\,\psi_{2n+1}^{(0)} \right\rangle = 0, 
\label{eq_mu1_2np1}\\
M\mu_{2n}^{(1)} & = & M\left\langle \psi_{2n}^{(0)}\Big|\,H_1\,\psi_{2n}^{(0)} \right\rangle = 0. 
\label{eq_mu1_2n}
\end{eqnarray}
This can be directly inferred from the well-defined symmetries of the zeroth-order eigenfunctions in Eqs.~(\ref{eq_0_2np1}) and (\ref{eq_0_2n}) and the uneven symmetry of $H_1$ in Eq.~(\ref{eq_H1}). 

In fact, we find along these lines that any integral of the kind
\begin{eqnarray}
\left\langle \psi_{2m+1}^{(0)}\Big|\,H_1\,\psi_{2n+1}^{(0)} \right\rangle &=& 0, 
\label{eq_sym_2np1}\\
\left\langle \psi_{2m}^{(0)}\Big|\,H_1\,\psi_{2n}^{(0)} \right\rangle &=& 0 
\label{eq_sym_2n}
\end{eqnarray}
vanishes. 
Apart from that, we introduce the abbreviation
\begin{equation}\label{eq_I}
I_{nm} := \int_{-\pi}^{\pi}\mathrm{d}v\,\cos\left(\frac{2n+1}{2}v\right)\sin\left(mv\right)\,\tan\left(\frac{v}{2}\right). 
\end{equation}
As a consequence, the first-order corrections to the eigenfunctions can be written as
\begin{widetext}
\begin{eqnarray}
M\psi_{2n+1}^{(1)} &=& 
M\sum_{m=1}^{\infty} \psi_{2m}^{(0)}\,\frac{\left\langle \psi_{2m}^{(0)}\Big|\,H_1\,\psi_{2n+1}^{(0)} \right\rangle}{\mu_{2n+1}^{(0)}-\mu_{2m}^{(0)}}
\;=\;
{}-\frac{M}{2\sqrt{\pi}^3}\sum_{m=1}^{\infty} \frac{ I_{nm} }{n(n+1)-m^2+\frac{1}{4}}\,\sin\left(mv\right), 
\label{eq_1_2np1}
\\[.2cm]
M\psi_{2n}^{(1)} &=& 
M\sum_{m=0}^{\infty} \psi_{2m+1}^{(0)}\,\frac{\left\langle \psi_{2m+1}^{(0)}\Big|\,H_1\,\psi_{2n}^{(0)} \right\rangle}{\mu_{2n}^{(0)}-\mu_{2m+1}^{(0)}}
\;=\;
{}-\frac{M}{2\sqrt{\pi}^3}\sum_{m=0}^{\infty} \frac{ I_{mn} }{n^2-\frac{1}{4}-m(m+1)}\,\cos\left(\frac{2m+1}{2}v\right). 
\label{eq_1_2n}
\end{eqnarray}

To second-order perturbation, we find the following corrections to the eigenvalues: 
%
%
%
\begin{eqnarray}
M^2\mu_{2n+1}^{(2)} &=& M^2\sum_{m=1}^{\infty}\frac{ 
\left\langle \psi_{2m}^{(0)}\Big|\,H_1\,\psi_{2n+1}^{(0)} \right\rangle 
^2}{\mu_{2n+1}^{(0)}-\mu_{2m}^{(0)}}
\;=\;
\frac{M^2}{4\pi^2}\sum_{m=1}^{\infty} \frac{ \left( I_{nm} \right)^2 }{n(n+1)-m^2+\frac{1}{4}}, 
\label{eq_mu2_2np1}
\\[.1cm]
M^2\mu_{2n}^{(2)} &=& 
M^2\sum_{m=0}^{\infty}\frac{ 
\left\langle \psi_{2m+1}^{(0)}\Big|\,H_1\,\psi_{2n}^{(0)} \right\rangle 
^2}{\mu_{2n}^{(0)}-\mu_{2m+1}^{(0)}}
\;=\;
\frac{M^2}{4\pi^2}\sum_{m=0}^{\infty} \frac{ \left( I_{mn} \right)^2 }{n^2-\frac{1}{4}-m(m+1)}. 
\label{eq_mu2_2n}
\end{eqnarray}
Taking into account Eqs.~(\ref{eq_sym_2np1}) and (\ref{eq_sym_2n}), the second-order corrections to the eigenfunctions are given by
\begin{eqnarray}
M^2\psi_{2n+1}^{(2)} &=& 
M^2\sum_{\substack{k=0\\[.05cm] k\neq n}}^{\infty}\psi_{2k+1}^{(0)}
\sum_{m=1}^{\infty}\frac{\left\langle \psi_{2k+1}^{(0)}\Big|\,H_1\,\psi_{2m}^{(0)} \right\rangle \left\langle \psi_{2m}^{(0)}\Big|\,H_1\,\psi_{2n+1}^{(0)} \right\rangle}{\left(\mu_{2n+1}^{(0)}-\mu_{2k+1}^{(0)}\right)\left(\mu_{2n+1}^{(0)}-\mu_{2m}^{(0)}\right)}
-\frac{M^2}{2}\psi_{2n+1}^{(0)}\sum_{k=1}^{\infty}\left( 
\frac{\left\langle \psi_{2n+1}^{(0)}\Big|\,H_1\,\psi_{2k}^{(0)} \right\rangle}{\mu_{2n+1}^{(0)}-\mu_{2k}^{(0)}}
\right)^2
\nonumber \\
&=&\frac{M^2}{4\pi^2\sqrt{\pi}}\,\sum_{\substack{k=0\\[.05cm] k\neq n}}^{\infty}\sum_{m=1}^{\infty}
\frac{ I_{km}\, I_{nm} }
{\left[n(n+1)-k(k+1)\right]\left[n(n+1)-m^2+\frac{1}{4}\right]}
\,\cos\left(\frac{2k+1}{2}v\right)
\nonumber\\
&&{}-\frac{M^2}{8\pi^2\sqrt{\pi}}\,
\sum_{k=1}^{\infty}\left( 
\frac{ I_{nk} }
{\left[n(n+1)-k^2+\frac{1}{4}\right]}
\right)^2\cos\left(\frac{2n+1}{2}v\right),
\label{eq_2_2np1}
\end{eqnarray}
\begin{eqnarray}
M^2\psi_{2n}^{(2)} &=&
M^2\sum_{\substack{k=1\\[.05cm] k\neq n}}^{\infty}\psi_{2k}^{(0)}
\sum_{m=0}^{\infty}\frac{\left\langle \psi_{2k}^{(0)}\Big|\,H_1\,\psi_{2m+1}^{(0)} \right\rangle \left\langle \psi_{2m+1}^{(0)}\Big|\,H_1\,\psi_{2n}^{(0)} \right\rangle}{\left(\mu_{2n}^{(0)}-\mu_{2k}^{(0)}\right)\left(\mu_{2n}^{(0)}-\mu_{2m+1}^{(0)}\right)}
-\frac{M^2}{2}\psi_{2n}^{(0)}\sum_{k=0}^{\infty}\left( 
\frac{\left\langle \psi_{2n}^{(0)}\Big|\,H_1\,\psi_{2k+1}^{(0)} \right\rangle}{\mu_{2n}^{(0)}-\mu_{2k+1}^{(0)}}
\right)^2
\nonumber \\
&=&\frac{M^2}{4\pi^2\sqrt{\pi}}\,\sum_{\substack{k=1\\[.05cm] k\neq n}}^{\infty}\sum_{m=0}^{\infty}
\frac{ I_{mk}\, I_{mn} }
{\left[n^2-k^2\right]\left[n^2-\frac{1}{4}-m(m+1)\right]}
\,\sin\left(kv\right)
\nonumber\\
&&{}-\frac{M^2}{8\pi^2\sqrt{\pi}}\,
\sum_{k=0}^{\infty}\left( 
\frac{ I_{kn} }
{\left[n^2-\frac{1}{4}-k(k+1)\right]}
\right)^2\sin\left(nv\right).
\label{eq_2_2n}
\end{eqnarray}
\pagebreak
\end{widetext}

In summary, from Eqs.~(\ref{eq_mu0_2np1}), (\ref{eq_mu0_2n}), (\ref{eq_mu1_2np1}), (\ref{eq_mu1_2n}), (\ref{eq_mu2_2np1}), and (\ref{eq_mu2_2n}) we have the perturbed eigenvalues
\begin{equation}
\mu_{2n+1} = n(n+1)+\frac{1}{4}M^2+M^2\mu_{2n+1}^{(2)},\quad  n=0,1,2,\dots,
\label{eq_mu_2np1}
\end{equation}
\begin{equation}
\mu_{2n} = n^2-\frac{1}{4}+\frac{1}{4}M^2+M^2\mu_{2n}^{(2)},\quad  n=1,2,\dots, 
\label{eq_mu_2n}
\end{equation}
while Eqs.~(\ref{eq_0_2np1}), (\ref{eq_0_2n}), (\ref{eq_psi2np1}), (\ref{eq_psi2n}), (\ref{eq_1_2np1}), (\ref{eq_1_2n}), (\ref{eq_2_2np1}), and (\ref{eq_2_2n}) 
lead us to the perturbed eigenfunctions. 
It is straightforward to verify that these perturbed eigenfunctions form again an orthonormal set. 

In analogy to Eqs.~(\ref{eq_expansion_v2np1}) and (\ref{eq_expansion_v2n}) as well as Eqs.~(\ref{eq_e01})--(\ref{eq_e04}), we expand the expressions $v\psi_{2n+1}$ and $v\psi_{2n}$ into our now perturbed orthonormal set and project onto $\psi_{2m+1}$ and $\psi_{2m}$, respectively. Up to second order in $M$, we obtain: 
\begin{widetext}
\begin{eqnarray}
\left\langle \psi_{2m+1}\Big|\,v\,\psi_{2n+1} \right\rangle 
&=&
M\left(
\left\langle \psi_{2m+1}^{(0)}\Big|\,v\,\psi_{2n+1}^{(1)} \right\rangle 
+\left\langle \psi_{2m+1}^{(1)}\Big|\,v\,\psi_{2n+1}^{(0)} \right\rangle 
\right)
\nonumber\\
&=&{}-\frac{M}{2\pi}
\sum_{k=1}^{\infty}\left(
\frac{I_{nk}\,J_{mk}}{n(n+1)-k^2+\frac{1}{4}}
+\frac{I_{mk}\,J_{nk}}{m(m+1)-k^2+\frac{1}{4}}
\right),
\label{eq_e1}\\[.3cm]
%
\left\langle \psi_{2m+1}\Big|\,v\,\psi_{2n} \right\rangle 
&=&
\left\langle \psi_{2m+1}^{(0)}\Big|\,v\,\psi_{2n}^{(0)} \right\rangle 
+M^2\left\langle \psi_{2m+1}^{(1)}\Big|\,v\,\psi_{2n}^{(1)} \right\rangle 
+M^2\left(
\left\langle \psi_{2m+1}^{(0)}\Big|\,v\,\psi_{2n}^{(2)} \right\rangle 
+\left\langle \psi_{2m+1}^{(2)}\Big|\,v\,\psi_{2n}^{(0)} \right\rangle 
\right)
\nonumber\\
&=&
J_{mn} + \frac{M^2}{4\pi^2}\sum_{k=1}^{\infty}\sum_{l=0}^{\infty}
\frac{I_{ln}\,I_{mk}}{\left[n^2-\frac{1}{4}-l(l+1)\right]\left[m(m+1)-k^2+\frac{1}{4}\right]}\,J_{lk}
\nonumber\\
&&{}+\frac{M^2}{4\pi^2}\Bigg(
\sum_{\substack{l=0\\[.05cm]l\neq m}}^{\infty}\sum_{k=1}^{\infty}
\frac{I_{lk}\,I_{mk}}{\left[m(m+1)-l(l+1)\right]\left[m(m+1)-k^2+\frac{1}{4}\right]}\,J_{ln}
\nonumber\\
&&{}\qquad\qquad\qquad+\sum_{\substack{l=1\\[.05cm]l\neq n}}^{\infty}\sum_{k=0}^{\infty}
\frac{I_{kl}\,I_{kn}}{\left[n^2-l^2\right]\left[n^2-\frac{1}{4}-k(k+1)\right]}\,J_{ml}
\Bigg)
\nonumber\\
&&{}-\frac{M^2}{8\pi^2}\,J_{mn}\left(
\sum_{l=1}^{\infty}\left(\frac{I_{ml}}{m(m+1)-l^2+\frac{1}{4}}\right)^2
+\sum_{l=0}^{\infty}\left(\frac{I_{ln}}{n^2-\frac{1}{4}-l(l+1)}\right)^2
\right)\\[.3cm]
%
\left\langle \psi_{2m}\Big|\,v\,\psi_{2n+1} \right\rangle 
&=&
\left\langle \psi_{2n+1}\Big|\,v\,\psi_{2m} \right\rangle \\[.3cm]
%
\left\langle \psi_{2m}\Big|\,v\,\psi_{2n} \right\rangle 
&=&
M\left(
\left\langle \psi_{2m}^{(0)}\Big|\,v\,\psi_{2n}^{(1)} \right\rangle 
+\left\langle \psi_{2m}^{(1)}\Big|\,v\,\psi_{2n}^{(0)} \right\rangle 
\right)
\nonumber\\
&=&{}-\frac{M}{2\pi}\sum_{k=0}^{\infty}\left(
\frac{I_{km}\,J_{kn}}{m^2-\frac{1}{4}-k(k+1)}
+\frac{I_{kn}\,J_{km}}{n^2-\frac{1}{4}-k(k+1)}
\right).
\label{eq_e4}
\end{eqnarray}
\end{widetext}

At the end of this whole procedure, we again obtain a system of equations as in Eq.~(\ref{eq_matrix}) for the expansion coefficients $\mathbf{c}(x,t)$. As before, $c_0(x,t)$ corresponds to the searched-for spatial distribution function. Furthermore, it is straightforward to verify that $\sqrt{f_{\mathrm{st}}(v)}=\psi_{2\cdot0+1}(v)$. Thus, from Eq.~(\ref{eq_init}), we again find the initial condition described by Eq.~(\ref{eq_c0init}). The entries of the matrix $\mathbf{e}$ are set in analogy to Eqs.~(\ref{eq_e01})--(\ref{eq_e04}), now, however, with the right-hand sides of 
these equations replaced by the corresponding expressions up to second order in $M$ listed in Eqs.~(\ref{eq_e1})--(\ref{eq_e4}). Unfortunately, the integrals in Eq.~(\ref{eq_I}) cannot be determined analytically, so we numerically evaluate and tabulate their magnitudes. The infinite sums in Eqs.~(\ref{eq_e1})--(\ref{eq_e4}) are cut at a certain index; we chose a cut-off value of $100$, which includes $201$ eigenfunctions in the calculation of the entries of the matrix $\mathbf{e}$. 
As before, the benefit of this whole procedure is that the numerical calculation can be significantly sped up. Cutting the system of equations for the expansion coefficients $\mathbf{c}$ in Eqs.~(\ref{eq_c2mp1}) and (\ref{eq_c2m}) at order $m=5$ is sufficient for our purposes; we found that the results do not noticeably change by including higher-order expansion coefficients. 


We have indicated our results obtained in this way for the spatial distribution function $c_0(x,t)$ in Fig.~\ref{fig_timeseries_Mneq0} and for the resulting moments of $x$ in Figs.~\ref{fig_drift_Mneq0}--\ref{fig_kurt_Mneq0}, respectively. They are marked by ``(sa)'' in the figures. For lower magnitudes of the drift force, here $M=0.2$, we find good agreement with the direct numerical solution of the corresponding Fokker-Planck equation, Eqs.~(\ref{eq_fp}) and (\ref{eq_init}). Since our approach is based on a perturbational expansion, the agreement naturally decreases with increasing magnitude of $M$. Remarkably, for $M=1$, we still observe reasonable qualitative representation of the numerically obtained data. For strong drift forces, here $M=5$, the perturbational approach breaks down; therefore, we did not show any corresponding data.

\section{Summary and conclusions}\label{sec_conclusions}

In summary, we investigated the velocity and displacement statistics resulting from a model of stochastic particle motion under nonlinear friction. We only considered particles that do not interact with each other. At low particle speeds, our friction grows approximately linearly with the particle velocity as for regular viscous friction. It grows stronger than linearly at higher speeds, while it diverges at a certain maximum in the possible particle speed. Starting from according Langevin equations, we derived the corresponding Fokker-Planck equation for one-dimensional motion. Moreover, we included the influence of a constant drift force. 

The resulting velocity distributions were calculated analytically, while the displacement statistics were obtained by particle-based simulations and, mainly, from solving the Fokker-Planck equation numerically. As a benefit of the model, a particular choice of the maximum particle speed ($a=\pi$ in rescaled units) allows significant analytical progress. In this way, the two-dimensional problem of solving the Fokker-Planck equation was reduced to a system of dynamic equations in the spatial coordinate only. Moreover, the resulting system of equations could be truncated at relatively low order. This procedure allows a significant speed-up in the overall calculation of the displacement statistics. The influence of the external drift force was included via perturbational expansion. 

In both cases, i.e.\ with and without an imposed drift force, we analyzed the resulting displacement statistics by addressing the time behavior of its first four moments. After an initial transient had decayed, the variance grew linearly in time as for regular diffusion under linear viscous friction. Nevertheless, the higher moments (via skewness and kurtosis) highlighted intermediate intervals of pronounced non-Gaussian displacement statistics. 

As a possible experimental realization of the model, we are thinking of the stochastic motion of colloidal particles in a complex fluid environment. If magnetic colloidal particles are used \cite{thurm2002magnetic,odenbach2003ferrofluids,huke2004magnetic,holm2005structure,klapp2005dipolar, alexiou2006targeting,tietze2013efficient,zaloga2014development,matuszak2015endothelial}, a constant drift force could be imposed by spatially homogeneous external magnetic field gradients. 
The statistics of particle displacements during diffusion could be obtained via single-particle tracking \cite{mason1997particle,anthony2006methods,saxton2008single, wang2009anomalous,roth2012simultaneous,huang2015buckling}. 
An example of appropriate complex fluid environments could be shear-thickening suspensions such as starch solutions \cite{fall2008shear,brown2014shear}. 
We recall that a memory term reflecting elastic aspects in the response of the surrounding medium has not yet been included into our model. Thus memory effects are not covered at the present stage of our description. 

In closing, we add some more technical remarks. First, we recall that significant analytical progress could only be achieved close to one special value of the maximum particle speed ($a=\pi$ in rescaled units). It might be possible to obtain approximative analytical solutions also for other values 
by using yet another perturbative expansion. 
\rev{In contrast to that, for linear viscous friction, the complete problem can be solved analytically \cite{risken1996fokker,zwanzig2001nonequilibrium}. However, the corresponding procedure does not carry over to nonlinear friction. This becomes plausible, for instance, by recognizing the central role that the Fourier transform plays in corresponding derivations \cite{risken1996fokker,zwanzig2001nonequilibrium}. In the case of Coulomb friction, a path integral approach provided further interesting insight into the dynamics and velocity statistics \cite{baule2010path,baule2011stick,chen2013weak}. Yet this formalism would need to be connected to the displacement statistics. Apart from that, related investigations were performed using the backward Fokker-Planck technique \cite{majumdar2002local}. An advantage of this method is that approximative analytical expressions for the displacement statistics are obtained \cite{chen2014large}. It was applied for the piecewise linear Coulomb friction model \cite{chen2014large} and might be extended to address nonlinear processes, possibly by approximating them by piecewise linear models.} 
Finally, one could think of including a memory term into the model to cover elastic parts in the environmental response. 
\rev{Instead of unbounded motion of a colloidal particle, confinement for example in a harmonic potential can be considered; some initial remarks on this point are summarized in Appendix~\ref{appendix_harmonic}.} 
In addition to that, the model could be extended to account for interactions between the particles in less diluted systems. A further natural step is to extend our description to more than one spatial dimension.

\begin{acknowledgments}
The author thanks Shilin Huang, G\"unter Auernhammer, Ralf Friedrich, Stefan Lyer, and Christoph Alexiou for stimulating discussions, as well as the Deutsche Forschungsgemeinschaft for support of this work through the priority program SPP 1681. 
\end{acknowledgments}

\appendix

\section{Comparison to the case of linear (viscous) friction}
\label{appendix_gauss}

\rev{
In the following, we further compare the results found in Sec.~\ref{barediffnonlin} to those obtained for the corresponding case of linear (viscous) friction. For this purpose, we expand the friction force in Eq.~(\ref{eq_friction}) to linear order, leading us to
\begin{equation}\label{eq_friction_lin}
F_{\mathrm{fr,lin}}(v) = -A\,\frac{\pi v}{2a}. 
\end{equation}
This is the same result as in Eq.~(\ref{eq_linearized_friction}) but now considered at all particle speeds $|v|$. 
The rescaled Fokker-Planck equation then reads
\begin{equation}\label{eq_fp_lin}
\partial_t f = {}-v\,\partial_x f + \partial_v\frac{\pi v}{2a}\,f  + \partial_v^2 f. 
\end{equation}
}

\rev{
In this case, the stationary velocity distribution is of Gaussian form, 
\begin{equation}\label{eq_fst_lin}
f_{\mathrm{st,lin}}(v) = \frac{1}{2\sqrt{a}}\,\exp\left\{ -\frac{\pi}{4a}v^2 \right\}. 
\end{equation}
Fig.~\ref{fig_vglgauss_fst} compares this Gaussian form to the results obtained from our nonlinear friction model in Eq.~(\ref{eq_fst}) and in Fig.~\ref{fig_fst_Meq0}. 
\begin{figure}
\centerline{\includegraphics[width=7.5cm]{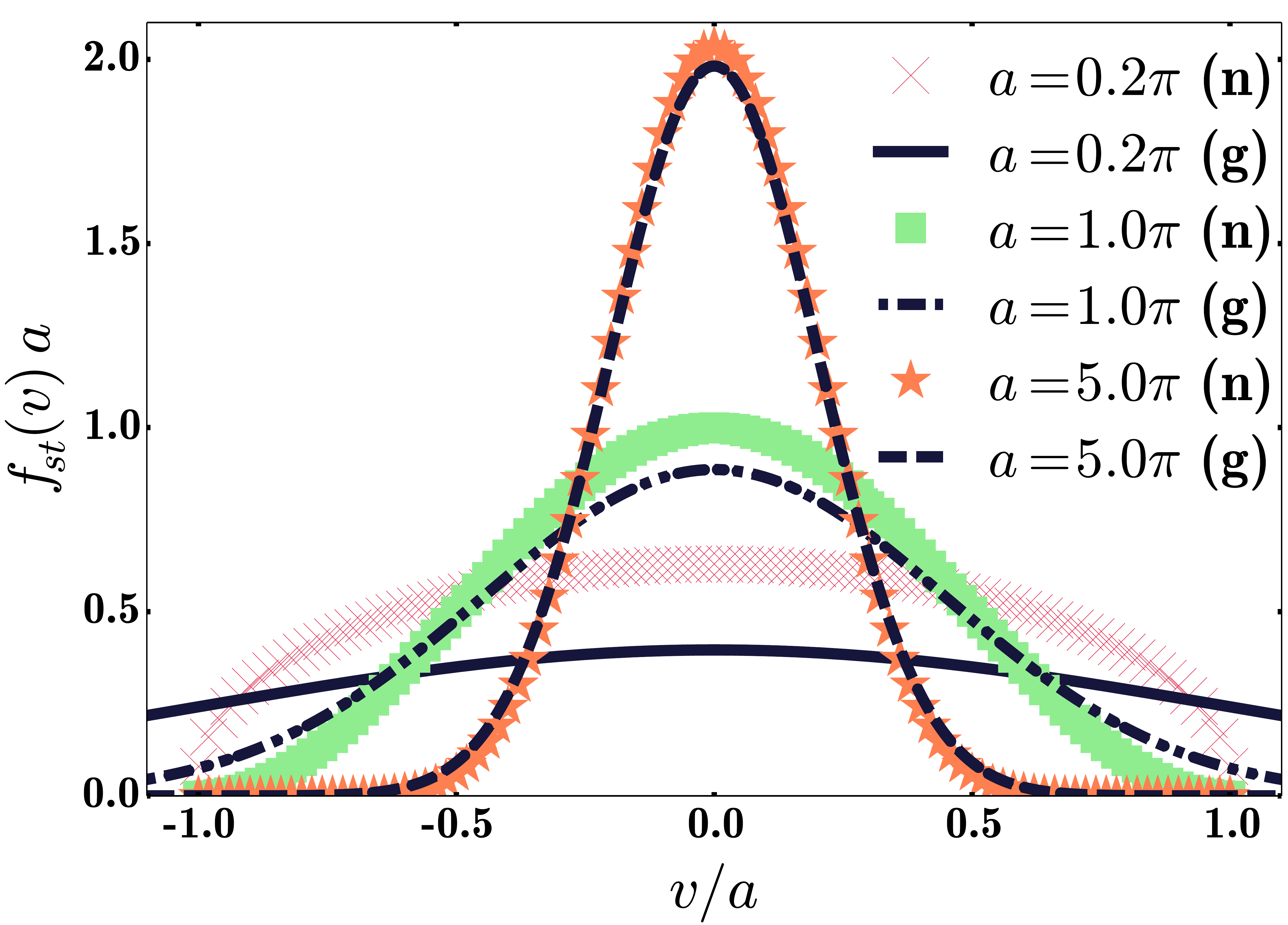}}
\caption{\rev{(Color online) Comparison between $f_{\mathrm{st}}(v)$ for the nonlinear friction model and $f_{\mathrm{st,lin}}(v)$ from Eq.~(\ref{eq_fst_lin}) for the linear expansion of the friction force. In the case of nonlinear friction, we plot again our numerical results from Fig.~\ref{fig_fst_Meq0} marked by ``(n)''. The Gaussian curves resulting for $f_{\mathrm{st,lin}}(v)$ are marked by ``(g)''. There is no drift force, $M=0$, and different magnitudes of the parameter $a$ are considered. For $a=0.2\pi$, the distributions in the nonlinear and linear case strongly differ from each other, while they become similar for increasing magnitude of $a$.} }
\label{fig_vglgauss_fst}
\end{figure}
For low magnitudes of $a$, here $a=0.2\pi$, the stationary velocity distributions in the nonlinear and linear friction models are markedly different from each other. With increasing magnitude of $a$, they become more and more similar. This can easily be understood by recognizing that the expression for the nonlinear friction force in Eq.~(\ref{eq_friction}) more and more approaches the linear expansion in Eq.~(\ref{eq_friction_lin}) with increasing magnitude of $a$.  
}

\rev{
Eq.~(\ref{eq_fp_lin}) can be solved analytically \cite{risken1996fokker,zwanzig2001nonequilibrium}. The resulting propagator, or transition probability density, is not explicitly reproduced here. It is of Gaussian functional form. After convolution with the initial condition Eq.~(\ref{eq_init}), using $f_{\mathrm{st,lin}}(v)$ from Eq.~(\ref{eq_fst_lin}) instead of $f_{\mathrm{st}}(v)$, and after integrating out the velocity variable, we obtain an analytical expression for the time dependent spatial distribution function under linear friction. It is again of Gaussian form. 
}

\rev{
For $a=\pi$, Fig.~\ref{fig_timeseries_vglgauss} compares the time evolution of the spatial distribution function under nonlinear friction, see Fig.~\ref{fig_timeseries_Meq0}, to the analytically calculated Gaussian one under linear friction. 
\begin{figure}
\centerline{\includegraphics[width=7.5cm]{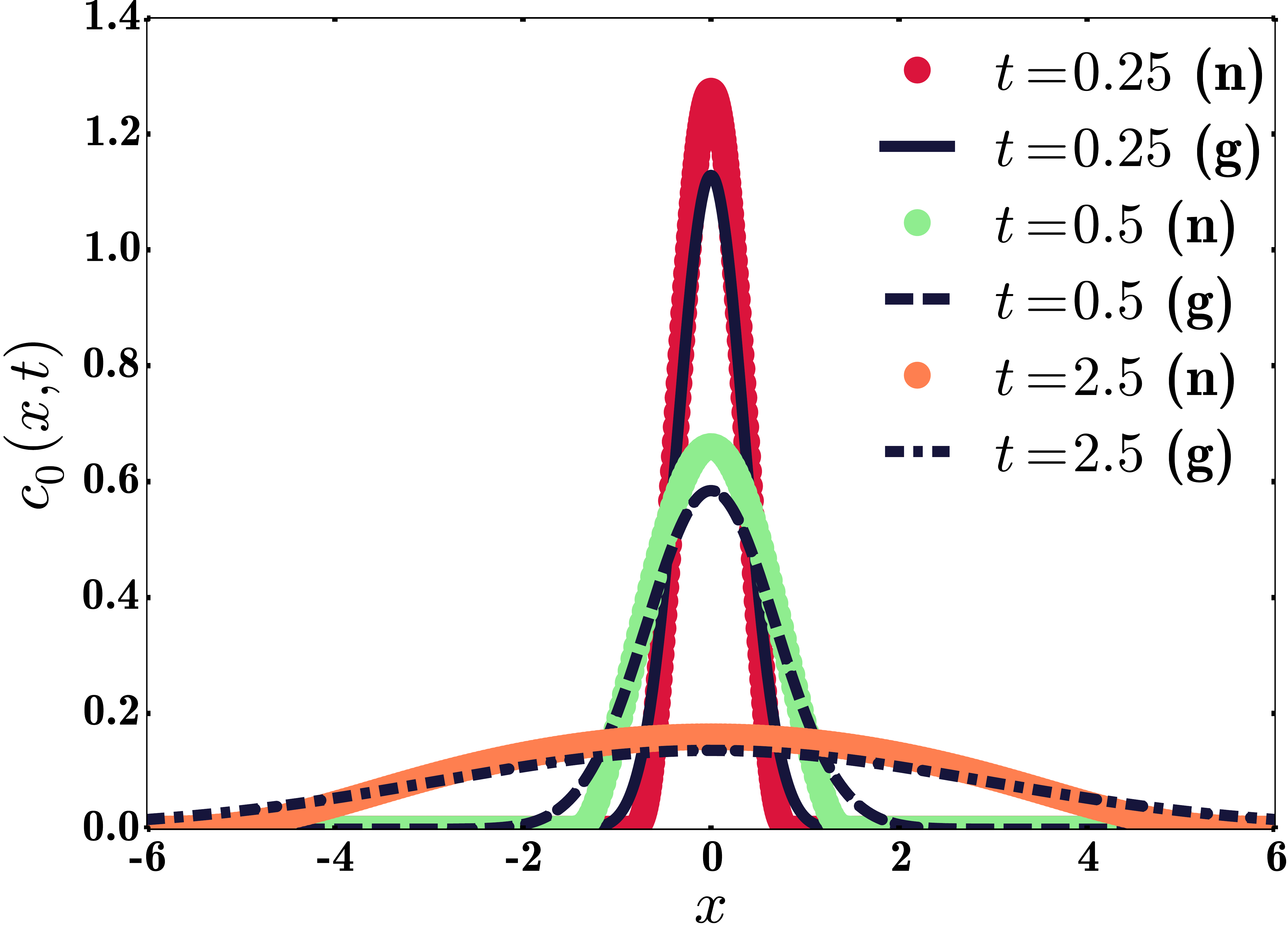}}
\caption{\rev{(Color online) Comparison between the time evolution of the spatial distribution function $c_0(x,t)$ under nonlinear friction and under linear friction. In the case of nonlinear friction, the results are obtained from direct numerical integration of the Fokker-Planck equation Eq.~(\ref{eq_fp}), marked by ``(n)'', and the same as in Fig.~\ref{fig_timeseries_Meq0}. The analytically calculated Gaussian curves in the case of linear friction are marked by ``(g)''. Moreover, $a=\pi$ and $M=0$.} }
\label{fig_timeseries_vglgauss}
\end{figure}
\begin{figure}
\centerline{\includegraphics[width=7.5cm]{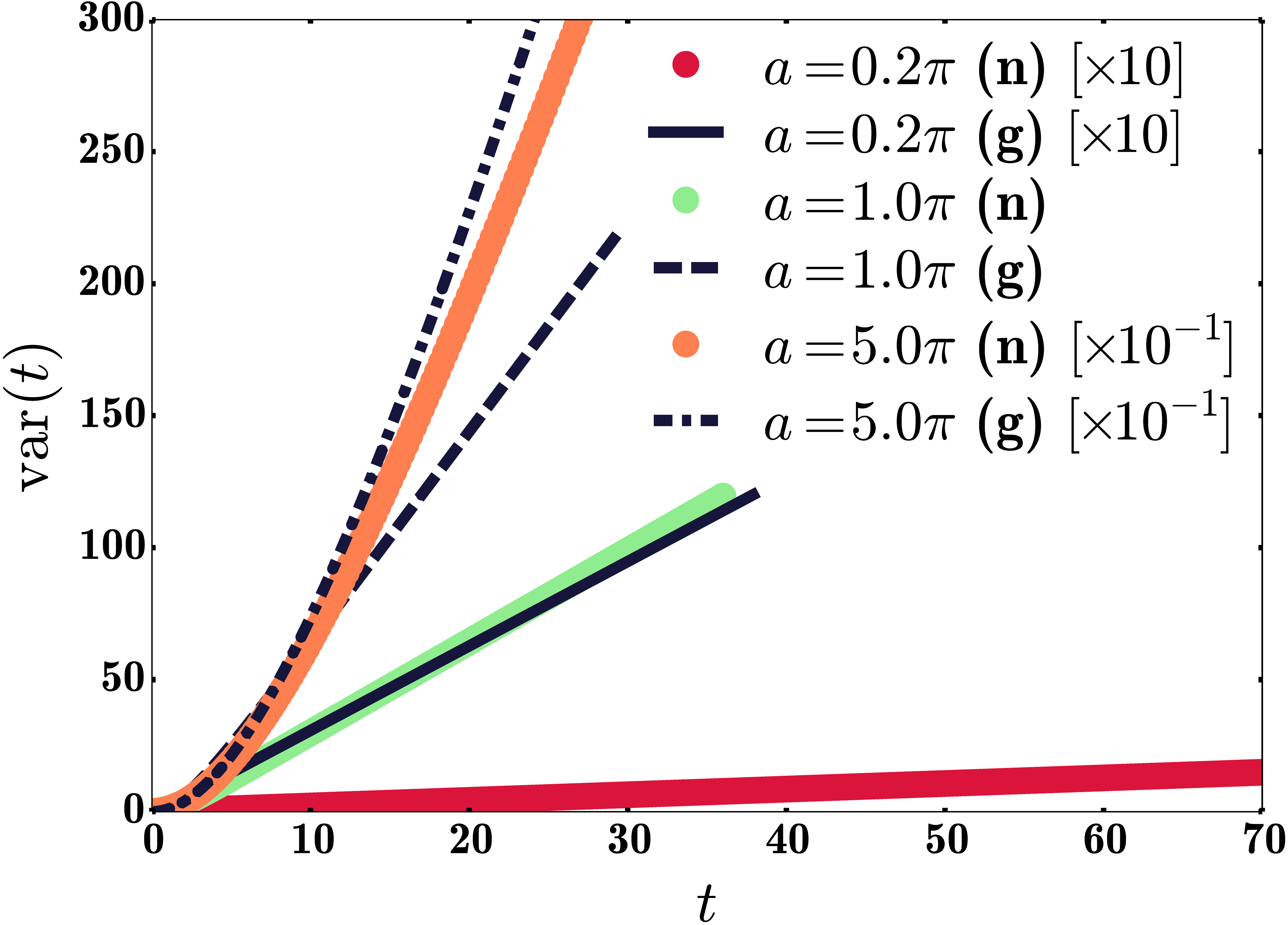}}
\caption{\rev{(Color online) Variance $\mathrm{var}(t)$ as a function of time $t$ under nonlinear friction and under linear friction. For nonlinear friction, the results are obtained numerically and marked by ``(n)''. They are the same as in Fig.~\ref{fig_var}. In the case of linear friction, the variances were calculated analytically from the obtained Gaussian curves and are marked by ``(g)''. We consider vanishing drift force $M=0$ and different values of the parameter $a$. Since $\langle x\rangle(t)=0$, the variances here coincide with the mean squared displacements. Data for $a=0.2\pi$ and $a=5.0\pi$ are multiplied by factors $10$ and $10^{-1}$, respectively, for better visualization.} }
\label{fig_var_vglgauss}
\end{figure}
\begin{figure}
\centerline{\includegraphics[width=7.5cm]{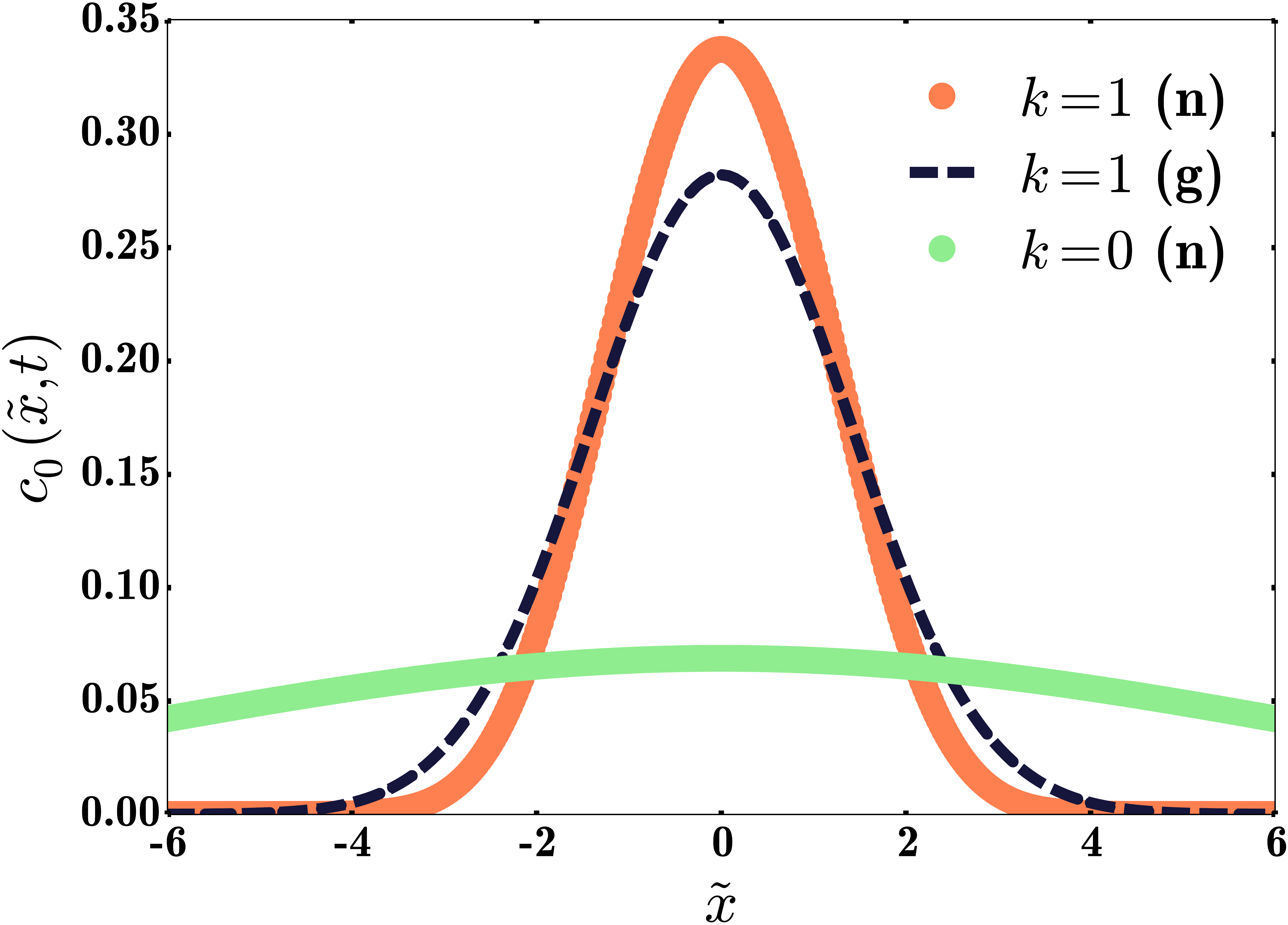}}
\caption{\rev{(Color online) Spatial distribution function $c_0(\tilde{x},t)$ under nonlinear friction, marked by ``(n)'', and under linear friction, marked by ``(g)'', under confinement. For nonlinear friction, the results are obtained from direct numerical integration of the Fokker-Planck equation Eq.~(\ref{eq_fp_confinement}) starting from the initial condition given by Eq.~(\ref{eq_init}). Under confinement, $k=1$, a steady state is attained, whereas the distribution function continues flattening without confinement, $k=0$. In the case of linear friction, an analytical solution exists for the steady state, see Eq.~(\ref{eq_c0_steady_gauss}), which is of Gaussian shape. It is less strongly peaked and less concentrated in the center of the confinement than the solution under nonlinear friction. The friction parameter is set to $a=\pi$ and the numerical curves are obtained at time $t=10$.} }
\label{fig_confinement}
\end{figure}
At identical time, the spatial distribution curve under linear friction is broader than under nonlinear friction. This is conceivable as our nonlinear friction bounds the magnitude of the particle velocity, which counteracts a rapid decay of the initially peaked spatial distribution function. 
}

\rev{
Furthermore, we compare the moments of the spatial distribution functions for nonlinear and linear friction. In both cases, since here we do not consider an imposed drift force, i.e.\ $M=0$, the odd moments vanish. 
Thus the variances in both cases are identical with the mean squared displacements. The variances are compared to each other in Fig.~\ref{fig_var_vglgauss}. 
There, one finds a situation similar to Fig.~\ref{fig_vglgauss_fst}. At low magnitude of $a$, the variances for both types of friction markedly differ from each other. In contrast to that, they become similar at higher magnitude of $a$. 
Furthermore, the kurtosis under nonlinear friction was plotted in Fig.~\ref{fig_kurtosis}. Under linear friction, the spatial distribution functions are of Gaussian shape. In this case, and together with the Gaussian initial conditions, the kurtosis vanishes at all times. 
}

\section{Confining harmonic potential}\label{appendix_harmonic}

\rev{
In addition to the nonlinear friction force in Eq.~(\ref{eq_friction}) and the constant drift force $M$, we now briefly consider the influence of a confining harmonic potential 
\begin{equation}\label{eq_U}
U(x) = \frac{1}{2}kx^2. 
\end{equation}
The constant $k>0$ sets the strength of the confinement. Again rescaling all quantities as described between Eqs.~(\ref{eq_fp_unscaled}) and (\ref{eq_fp}), defining $k=[mA^4/(Kk_BT)^2]k'$, and afterwards omitting the primes, the corresponding Fokker-Planck equation reads
\begin{eqnarray}
\partial_t f &=& {}-v\,\partial_x f + \partial_v \tan\left(\frac{\pi v}{2a}\right)f - M \,\partial_v f \nonumber\\[.1cm]
&&{}+ kx\,\partial_v f+ \partial_v^2 f. \\[-.3cm] \nonumber
\end{eqnarray}
As an advantage, the influence of the drift force $M$ is now readily and exactly included by a simple coordinate transform $x=\tilde{x}+M/k$. Then the Fokker-Planck equation becomes
\begin{equation}\label{eq_fp_confinement}
\partial_t f = {}-v\,\partial_{\tilde{x}} f + \partial_v \tan\left(\frac{\pi v}{2a}\right)f  + k\tilde{x}\,\partial_v f+ \partial_v^2 f. 
\end{equation}
}

\rev{
Due to the confinement, starting from the initial condition Eq.~(\ref{eq_init}), the spatial distribution function does not continuously flatten any more as in Fig.~\ref{fig_timeseries_Meq0}. Instead, it attains a steady state after an intermediate transient time. An example is shown in Fig.~\ref{fig_confinement}. 
}

\rev{
There, we numerically iterate Eq.~(\ref{eq_fp_confinement}) with initial condition Eq.~(\ref{eq_init}) forward in time until the steady state is reached. In contrast to that, without confinement, i.e.\ for $k=0$, the spatial distribution function keeps on flattening, as indicated in the figure. We compare the situation under nonlinear friction as given by Eq.~(\ref{eq_friction}) to the case of linear (viscous) friction as formulated in Eq.~(\ref{eq_friction_lin}). In the nonlinear case, the distribution function is peaked more strongly. This is because higher velocity magnitudes are less likely or completely prohibited, see Fig.~\ref{fig_fst_Meq0}. Therefore, a particle under nonlinear friction cannot obtain as much drive as a quick particle under linear friction. Thus the spatial probability distribution cannot work as strongly against the confining potential and remains more concentrated around the center. The linear case can be solved analytically \cite{risken1996fokker} and leads to a steady spatial distribution function of Gaussian shape, 
\begin{equation}\label{eq_c0_steady_gauss}
c_{0,\mathrm{lin}}(\tilde{x},t\rightarrow\infty) = 
\frac{1}{2}\sqrt{\frac{k}{a}} \exp\left\{ {}-k\,\frac{\pi}{4a}\,\tilde{x}^2 \right\}
\end{equation}
as indicated in Fig.~\ref{fig_confinement}.
}


\begin{thebibliography}{59}%
\makeatletter
\providecommand \@ifxundefined [1]{%
 \@ifx{#1\undefined}
}%
\providecommand \@ifnum [1]{%
 \ifnum #1\expandafter \@firstoftwo
 \else \expandafter \@secondoftwo
 \fi
}%
\providecommand \@ifx [1]{%
 \ifx #1\expandafter \@firstoftwo
 \else \expandafter \@secondoftwo
 \fi
}%
\providecommand \natexlab [1]{#1}%
\providecommand \enquote  [1]{``#1''}%
\providecommand \bibnamefont  [1]{#1}%
\providecommand \bibfnamefont [1]{#1}%
\providecommand \citenamefont [1]{#1}%
\providecommand \href@noop [0]{\@secondoftwo}%
\providecommand \href [0]{\begingroup \@sanitize@url \@href}%
\providecommand \@href[1]{\@@startlink{#1}\@@href}%
\providecommand \@@href[1]{\endgroup#1\@@endlink}%
\providecommand \@sanitize@url [0]{\catcode `\\12\catcode `\$12\catcode
  `\&12\catcode `\#12\catcode `\^12\catcode `\_12\catcode `\%12\relax}%
\providecommand \@@startlink[1]{}%
\providecommand \@@endlink[0]{}%
\providecommand \url  [0]{\begingroup\@sanitize@url \@url }%
\providecommand \@url [1]{\endgroup\@href {#1}{\urlprefix }}%
\providecommand \urlprefix  [0]{URL }%
\providecommand \Eprint [0]{\href }%
\providecommand \doibase [0]{http://dx.doi.org/}%
\providecommand \selectlanguage [0]{\@gobble}%
\providecommand \bibinfo  [0]{\@secondoftwo}%
\providecommand \bibfield  [0]{\@secondoftwo}%
\providecommand \translation [1]{[#1]}%
\providecommand \BibitemOpen [0]{}%
\providecommand \bibitemStop [0]{}%
\providecommand \bibitemNoStop [0]{.\EOS\space}%
\providecommand \EOS [0]{\spacefactor3000\relax}%
\providecommand \BibitemShut  [1]{\csname bibitem#1\endcsname}%
\let\auto@bib@innerbib\@empty
\bibitem [{\citenamefont {Wang}\ \emph {et~al.}(2009)\citenamefont {Wang},
  \citenamefont {Anthony}, \citenamefont {Bae},\ and\ \citenamefont
  {Granick}}]{wang2009anomalous}%
  \BibitemOpen
  \bibfield  {author} {\bibinfo {author} {\bibfnamefont {B.}~\bibnamefont
  {Wang}}, \bibinfo {author} {\bibfnamefont {S.~M.}\ \bibnamefont {Anthony}},
  \bibinfo {author} {\bibfnamefont {S.~C.}\ \bibnamefont {Bae}}, \ and\
  \bibinfo {author} {\bibfnamefont {S.}~\bibnamefont {Granick}},\ }\href@noop
  {} {\bibfield  {journal} {\bibinfo  {journal} {Proc. Natl. Acad. Sci. USA}\
  }\textbf {\bibinfo {volume} {106}},\ \bibinfo {pages} {15160} (\bibinfo
  {year} {2009})}\BibitemShut {NoStop}%
\bibitem [{\citenamefont {Wang}\ \emph {et~al.}(2012)\citenamefont {Wang},
  \citenamefont {Kuo}, \citenamefont {Bae},\ and\ \citenamefont
  {Granick}}]{wang2012brownian}%
  \BibitemOpen
  \bibfield  {author} {\bibinfo {author} {\bibfnamefont {B.}~\bibnamefont
  {Wang}}, \bibinfo {author} {\bibfnamefont {J.}~\bibnamefont {Kuo}}, \bibinfo
  {author} {\bibfnamefont {S.~C.}\ \bibnamefont {Bae}}, \ and\ \bibinfo
  {author} {\bibfnamefont {S.}~\bibnamefont {Granick}},\ }\href@noop {}
  {\bibfield  {journal} {\bibinfo  {journal} {Nature Mater.}\ }\textbf
  {\bibinfo {volume} {11}},\ \bibinfo {pages} {481} (\bibinfo {year}
  {2012})}\BibitemShut {NoStop}%
\bibitem [{\citenamefont {H{\"o}fling}\ and\ \citenamefont
  {Franosch}(2013)}]{hofling2013anomalous}%
  \BibitemOpen
  \bibfield  {author} {\bibinfo {author} {\bibfnamefont {F.}~\bibnamefont
  {H{\"o}fling}}\ and\ \bibinfo {author} {\bibfnamefont {T.}~\bibnamefont
  {Franosch}},\ }\href@noop {} {\bibfield  {journal} {\bibinfo  {journal} {Rep.
  Prog. Phys.}\ }\textbf {\bibinfo {volume} {76}},\ \bibinfo {pages} {046602}
  (\bibinfo {year} {2013})}\BibitemShut {NoStop}%
\bibitem [{\citenamefont {Guan}\ \emph {et~al.}(2014)\citenamefont {Guan},
  \citenamefont {Wang},\ and\ \citenamefont {Granick}}]{guan2014even}%
  \BibitemOpen
  \bibfield  {author} {\bibinfo {author} {\bibfnamefont {J.}~\bibnamefont
  {Guan}}, \bibinfo {author} {\bibfnamefont {B.}~\bibnamefont {Wang}}, \ and\
  \bibinfo {author} {\bibfnamefont {S.}~\bibnamefont {Granick}},\ }\href@noop
  {} {\bibfield  {journal} {\bibinfo  {journal} {ACS Nano}\ }\textbf {\bibinfo
  {volume} {8}},\ \bibinfo {pages} {3331} (\bibinfo {year} {2014})}\BibitemShut
  {NoStop}%
\bibitem [{\citenamefont {Kalathi}\ \emph {et~al.}(2014)\citenamefont
  {Kalathi}, \citenamefont {Yamamoto}, \citenamefont {Schweizer}, \citenamefont
  {Grest},\ and\ \citenamefont {Kumar}}]{kalathi2014nanoparticle}%
  \BibitemOpen
  \bibfield  {author} {\bibinfo {author} {\bibfnamefont {J.~T.}\ \bibnamefont
  {Kalathi}}, \bibinfo {author} {\bibfnamefont {U.}~\bibnamefont {Yamamoto}},
  \bibinfo {author} {\bibfnamefont {K.~S.}\ \bibnamefont {Schweizer}}, \bibinfo
  {author} {\bibfnamefont {G.~S.}\ \bibnamefont {Grest}}, \ and\ \bibinfo
  {author} {\bibfnamefont {S.~K.}\ \bibnamefont {Kumar}},\ }\href@noop {}
  {\bibfield  {journal} {\bibinfo  {journal} {Phys. Rev. Lett.}\ }\textbf
  {\bibinfo {volume} {112}},\ \bibinfo {pages} {108301} (\bibinfo {year}
  {2014})}\BibitemShut {NoStop}%
\bibitem [{\citenamefont {Babaye~Khorasani}\ \emph {et~al.}(2014)\citenamefont
  {Babaye~Khorasani}, \citenamefont {Poling-Skutvik}, \citenamefont
  {Krishnamoorti},\ and\ \citenamefont {Conrad}}]{babaye2014mobility}%
  \BibitemOpen
  \bibfield  {author} {\bibinfo {author} {\bibfnamefont {F.}~\bibnamefont
  {Babaye~Khorasani}}, \bibinfo {author} {\bibfnamefont {R.}~\bibnamefont
  {Poling-Skutvik}}, \bibinfo {author} {\bibfnamefont {R.}~\bibnamefont
  {Krishnamoorti}}, \ and\ \bibinfo {author} {\bibfnamefont {J.~C.}\
  \bibnamefont {Conrad}},\ }\href@noop {} {\bibfield  {journal} {\bibinfo
  {journal} {Macromolecules}\ }\textbf {\bibinfo {volume} {47}},\ \bibinfo
  {pages} {5328} (\bibinfo {year} {2014})}\BibitemShut {NoStop}%
\bibitem [{\citenamefont {Kubo}\ \emph {et~al.}(1991)\citenamefont {Kubo},
  \citenamefont {Toda},\ and\ \citenamefont
  {Hashitsume}}]{kubo1991statistical}%
  \BibitemOpen
  \bibfield  {author} {\bibinfo {author} {\bibfnamefont {R.}~\bibnamefont
  {Kubo}}, \bibinfo {author} {\bibfnamefont {M.}~\bibnamefont {Toda}}, \ and\
  \bibinfo {author} {\bibfnamefont {N.}~\bibnamefont {Hashitsume}},\
  }\href@noop {} {\emph {\bibinfo {title} {Statistical Physics II}}}\ (\bibinfo
   {publisher} {Springer, Berlin},\ \bibinfo {year} {1991})\BibitemShut
  {NoStop}%
\bibitem [{\citenamefont {Zwanzig}(2001)}]{zwanzig2001nonequilibrium}%
  \BibitemOpen
  \bibfield  {author} {\bibinfo {author} {\bibfnamefont {R.}~\bibnamefont
  {Zwanzig}},\ }\href@noop {} {\emph {\bibinfo {title} {{Nonequilibrium
  Statistical Mechanics}}}}\ (\bibinfo  {publisher} {Oxford University Press,
  Oxford},\ \bibinfo {year} {2001})\BibitemShut {NoStop}%
\bibitem [{\citenamefont {van Kampen}(2007)}]{kampen2007stochastic}%
  \BibitemOpen
  \bibfield  {author} {\bibinfo {author} {\bibfnamefont {N.~G.}\ \bibnamefont
  {van Kampen}},\ }\href@noop {} {\emph {\bibinfo {title} {Stochastic Processes
  in Physics and Chemistry}}}\ (\bibinfo  {publisher} {Elsevier, Amsterdam},\
  \bibinfo {year} {2007})\BibitemShut {NoStop}%
\bibitem [{\citenamefont {Gardiner}(2009)}]{gardiner2009stochastic}%
  \BibitemOpen
  \bibfield  {author} {\bibinfo {author} {\bibfnamefont {C.}~\bibnamefont
  {Gardiner}},\ }\href@noop {} {\emph {\bibinfo {title} {Stochastic Methods: A
  Handbook for the Natural and Social Sciences}}}\ (\bibinfo  {publisher}
  {Springer, Berlin},\ \bibinfo {year} {2009})\BibitemShut {NoStop}%
\bibitem [{\citenamefont {Chubynsky}\ and\ \citenamefont
  {Slater}(2014)}]{chubynsky2014diffusing}%
  \BibitemOpen
  \bibfield  {author} {\bibinfo {author} {\bibfnamefont {M.~V.}\ \bibnamefont
  {Chubynsky}}\ and\ \bibinfo {author} {\bibfnamefont {G.~W.}\ \bibnamefont
  {Slater}},\ }\href@noop {} {\bibfield  {journal} {\bibinfo  {journal} {Phys.
  Rev. Lett.}\ }\textbf {\bibinfo {volume} {113}},\ \bibinfo {pages} {098302}
  (\bibinfo {year} {2014})}\BibitemShut {NoStop}%
\bibitem [{\citenamefont {Kegel}\ and\ \citenamefont {van
  Blaaderen}(2000)}]{kegel2000direct}%
  \BibitemOpen
  \bibfield  {author} {\bibinfo {author} {\bibfnamefont {W.~K.}\ \bibnamefont
  {Kegel}}\ and\ \bibinfo {author} {\bibfnamefont {A.}~\bibnamefont {van
  Blaaderen}},\ }\href@noop {} {\bibfield  {journal} {\bibinfo  {journal}
  {Science}\ }\textbf {\bibinfo {volume} {287}},\ \bibinfo {pages} {290}
  (\bibinfo {year} {2000})}\BibitemShut {NoStop}%
\bibitem [{\citenamefont {Zheng}\ \emph {et~al.}(2013)\citenamefont {Zheng},
  \citenamefont {ten Hagen}, \citenamefont {Kaiser}, \citenamefont {Wu},
  \citenamefont {Cui}, \citenamefont {Silber-Li},\ and\ \citenamefont
  {L{\"o}wen}}]{zheng2013non}%
  \BibitemOpen
  \bibfield  {author} {\bibinfo {author} {\bibfnamefont {X.}~\bibnamefont
  {Zheng}}, \bibinfo {author} {\bibfnamefont {B.}~\bibnamefont {ten Hagen}},
  \bibinfo {author} {\bibfnamefont {A.}~\bibnamefont {Kaiser}}, \bibinfo
  {author} {\bibfnamefont {M.}~\bibnamefont {Wu}}, \bibinfo {author}
  {\bibfnamefont {H.}~\bibnamefont {Cui}}, \bibinfo {author} {\bibfnamefont
  {Z.}~\bibnamefont {Silber-Li}}, \ and\ \bibinfo {author} {\bibfnamefont
  {H.}~\bibnamefont {L{\"o}wen}},\ }\href@noop {} {\bibfield  {journal}
  {\bibinfo  {journal} {Phys. Rev. E}\ }\textbf {\bibinfo {volume} {88}},\
  \bibinfo {pages} {032304} (\bibinfo {year} {2013})}\BibitemShut {NoStop}%
\bibitem [{\citenamefont {Goohpattader}\ \emph {et~al.}(2009)\citenamefont
  {Goohpattader}, \citenamefont {Mettu},\ and\ \citenamefont
  {Chaudhury}}]{goohpattader2009experimental}%
  \BibitemOpen
  \bibfield  {author} {\bibinfo {author} {\bibfnamefont {P.~S.}\ \bibnamefont
  {Goohpattader}}, \bibinfo {author} {\bibfnamefont {S.}~\bibnamefont {Mettu}},
  \ and\ \bibinfo {author} {\bibfnamefont {M.~K.}\ \bibnamefont {Chaudhury}},\
  }\href@noop {} {\bibfield  {journal} {\bibinfo  {journal} {Langmuir}\
  }\textbf {\bibinfo {volume} {25}},\ \bibinfo {pages} {9969} (\bibinfo {year}
  {2009})}\BibitemShut {NoStop}%
\bibitem [{\citenamefont {Goohpattader}\ and\ \citenamefont
  {Chaudhury}(2010)}]{goohpattader2010diffusive}%
  \BibitemOpen
  \bibfield  {author} {\bibinfo {author} {\bibfnamefont {P.~S.}\ \bibnamefont
  {Goohpattader}}\ and\ \bibinfo {author} {\bibfnamefont {M.~K.}\ \bibnamefont
  {Chaudhury}},\ }\href@noop {} {\bibfield  {journal} {\bibinfo  {journal} {J.
  Chem. Phys.}\ }\textbf {\bibinfo {volume} {133}},\ \bibinfo {pages} {024702}
  (\bibinfo {year} {2010})}\BibitemShut {NoStop}%
\bibitem [{\citenamefont {Goohpattader}\ \emph {et~al.}(2011)\citenamefont
  {Goohpattader}, \citenamefont {Mettu},\ and\ \citenamefont
  {Chaudhury}}]{goohpattader2011stochastic}%
  \BibitemOpen
  \bibfield  {author} {\bibinfo {author} {\bibfnamefont {P.~S.}\ \bibnamefont
  {Goohpattader}}, \bibinfo {author} {\bibfnamefont {S.}~\bibnamefont {Mettu}},
  \ and\ \bibinfo {author} {\bibfnamefont {M.~K.}\ \bibnamefont {Chaudhury}},\
  }\href@noop {} {\bibfield  {journal} {\bibinfo  {journal} {Eur. Phys. J. E}\
  }\textbf {\bibinfo {volume} {34}},\ \bibinfo {pages} {120} (\bibinfo {year}
  {2011})}\BibitemShut {NoStop}%
\bibitem [{\citenamefont {Persson}(2000)}]{persson2000sliding}%
  \BibitemOpen
  \bibfield  {author} {\bibinfo {author} {\bibfnamefont {B.~N.~J.}\
  \bibnamefont {Persson}},\ }\href@noop {} {\emph {\bibinfo {title} {Sliding
  Friction: Physical Principles and Applications}}}\ (\bibinfo  {publisher}
  {Springer},\ \bibinfo {year} {2000})\BibitemShut {NoStop}%
\bibitem [{\citenamefont {Kawarada}\ and\ \citenamefont
  {Hayakawa}(2004)}]{kawarada2004non}%
  \BibitemOpen
  \bibfield  {author} {\bibinfo {author} {\bibfnamefont {A.}~\bibnamefont
  {Kawarada}}\ and\ \bibinfo {author} {\bibfnamefont {H.}~\bibnamefont
  {Hayakawa}},\ }\href@noop {} {\bibfield  {journal} {\bibinfo  {journal} {J.
  Phys. Soc. Jpn.}\ }\textbf {\bibinfo {volume} {73}},\ \bibinfo {pages} {2037}
  (\bibinfo {year} {2004})}\BibitemShut {NoStop}%
\bibitem [{\citenamefont {de~Gennes}(2005)}]{gennes2005brownian}%
  \BibitemOpen
  \bibfield  {author} {\bibinfo {author} {\bibfnamefont {P.~G.}\ \bibnamefont
  {de~Gennes}},\ }\href@noop {} {\bibfield  {journal} {\bibinfo  {journal} {J.
  Stat. Phys.}\ }\textbf {\bibinfo {volume} {119}},\ \bibinfo {pages} {953}
  (\bibinfo {year} {2005})}\BibitemShut {NoStop}%
\bibitem [{\citenamefont {Hayakawa}(2005)}]{hayakawa2005langevin}%
  \BibitemOpen
  \bibfield  {author} {\bibinfo {author} {\bibfnamefont {H.}~\bibnamefont
  {Hayakawa}},\ }\href@noop {} {\bibfield  {journal} {\bibinfo  {journal}
  {Physica D}\ }\textbf {\bibinfo {volume} {205}},\ \bibinfo {pages} {48}
  (\bibinfo {year} {2005})}\BibitemShut {NoStop}%
\bibitem [{\citenamefont {Touchette}\ \emph {et~al.}(2010)\citenamefont
  {Touchette}, \citenamefont {Van~der Straeten},\ and\ \citenamefont
  {Just}}]{touchette2010brownian}%
  \BibitemOpen
  \bibfield  {author} {\bibinfo {author} {\bibfnamefont {H.}~\bibnamefont
  {Touchette}}, \bibinfo {author} {\bibfnamefont {E.}~\bibnamefont {Van~der
  Straeten}}, \ and\ \bibinfo {author} {\bibfnamefont {W.}~\bibnamefont
  {Just}},\ }\href@noop {} {\bibfield  {journal} {\bibinfo  {journal} {J. Phys.
  A: Math. Theor.}\ }\textbf {\bibinfo {volume} {43}},\ \bibinfo {pages}
  {445002} (\bibinfo {year} {2010})}\BibitemShut {NoStop}%
\bibitem [{\citenamefont {Baule}\ \emph {et~al.}(2011)\citenamefont {Baule},
  \citenamefont {Touchette},\ and\ \citenamefont {Cohen}}]{baule2011stick}%
  \BibitemOpen
  \bibfield  {author} {\bibinfo {author} {\bibfnamefont {A.}~\bibnamefont
  {Baule}}, \bibinfo {author} {\bibfnamefont {H.}~\bibnamefont {Touchette}}, \
  and\ \bibinfo {author} {\bibfnamefont {E.~G.~D.}\ \bibnamefont {Cohen}},\
  }\href@noop {} {\bibfield  {journal} {\bibinfo  {journal} {Nonlinearity}\
  }\textbf {\bibinfo {volume} {24}},\ \bibinfo {pages} {351} (\bibinfo {year}
  {2011})}\BibitemShut {NoStop}%
\bibitem [{\citenamefont {Menzel}\ and\ \citenamefont
  {Goldenfeld}(2011)}]{menzel2011effect}%
  \BibitemOpen
  \bibfield  {author} {\bibinfo {author} {\bibfnamefont {A.~M.}\ \bibnamefont
  {Menzel}}\ and\ \bibinfo {author} {\bibfnamefont {N.}~\bibnamefont
  {Goldenfeld}},\ }\href@noop {} {\bibfield  {journal} {\bibinfo  {journal}
  {Phys. Rev. E}\ }\textbf {\bibinfo {volume} {84}},\ \bibinfo {pages} {011122}
  (\bibinfo {year} {2011})}\BibitemShut {NoStop}%
\bibitem [{\citenamefont {Talbot}\ and\ \citenamefont
  {Viot}(2012)}]{talbot2012effect}%
  \BibitemOpen
  \bibfield  {author} {\bibinfo {author} {\bibfnamefont {J.}~\bibnamefont
  {Talbot}}\ and\ \bibinfo {author} {\bibfnamefont {P.}~\bibnamefont {Viot}},\
  }\href@noop {} {\bibfield  {journal} {\bibinfo  {journal} {Phys. Rev. E}\
  }\textbf {\bibinfo {volume} {85}},\ \bibinfo {pages} {021310} (\bibinfo
  {year} {2012})}\BibitemShut {NoStop}%
\bibitem [{\citenamefont {Baule}\ and\ \citenamefont
  {Sollich}(2012)}]{baule2012singular}%
  \BibitemOpen
  \bibfield  {author} {\bibinfo {author} {\bibfnamefont {A.}~\bibnamefont
  {Baule}}\ and\ \bibinfo {author} {\bibfnamefont {P.}~\bibnamefont
  {Sollich}},\ }\href@noop {} {\bibfield  {journal} {\bibinfo  {journal}
  {Europhys. Lett.}\ }\textbf {\bibinfo {volume} {97}},\ \bibinfo {pages}
  {20001} (\bibinfo {year} {2012})}\BibitemShut {NoStop}%
\bibitem [{\citenamefont {Baule}\ and\ \citenamefont
  {Sollich}(2013)}]{baule2013rectification}%
  \BibitemOpen
  \bibfield  {author} {\bibinfo {author} {\bibfnamefont {A.}~\bibnamefont
  {Baule}}\ and\ \bibinfo {author} {\bibfnamefont {P.}~\bibnamefont
  {Sollich}},\ }\href@noop {} {\bibfield  {journal} {\bibinfo  {journal} {Phys.
  Rev. E}\ }\textbf {\bibinfo {volume} {87}},\ \bibinfo {pages} {032112}
  (\bibinfo {year} {2013})}\BibitemShut {NoStop}%
\bibitem [{\citenamefont {Pototsky}\ and\ \citenamefont
  {Marchesoni}(2013)}]{pototsky2013periodically}%
  \BibitemOpen
  \bibfield  {author} {\bibinfo {author} {\bibfnamefont {A.}~\bibnamefont
  {Pototsky}}\ and\ \bibinfo {author} {\bibfnamefont {F.}~\bibnamefont
  {Marchesoni}},\ }\href@noop {} {\bibfield  {journal} {\bibinfo  {journal}
  {Phys. Rev. E}\ }\textbf {\bibinfo {volume} {87}},\ \bibinfo {pages} {032132}
  (\bibinfo {year} {2013})}\BibitemShut {NoStop}%
\bibitem [{\citenamefont {Kanazawa}\ \emph {et~al.}(2015)\citenamefont
  {Kanazawa}, \citenamefont {Sano}, \citenamefont {Sagawa},\ and\ \citenamefont
  {Hayakawa}}]{kanazawa2015asymptotic}%
  \BibitemOpen
  \bibfield  {author} {\bibinfo {author} {\bibfnamefont {K.}~\bibnamefont
  {Kanazawa}}, \bibinfo {author} {\bibfnamefont {T.~G.}\ \bibnamefont {Sano}},
  \bibinfo {author} {\bibfnamefont {T.}~\bibnamefont {Sagawa}}, \ and\ \bibinfo
  {author} {\bibfnamefont {H.}~\bibnamefont {Hayakawa}},\ }\href@noop {}
  {\bibfield  {journal} {\bibinfo  {journal} {J. Stat. Phys.}\ }\textbf
  {\bibinfo {volume} {160}},\ \bibinfo {pages} {1294} (\bibinfo {year}
  {2015})}\BibitemShut {NoStop}%
\bibitem [{\citenamefont {Sano}\ and\ \citenamefont
  {Hayakawa}(2014)}]{sano2014roles}%
  \BibitemOpen
  \bibfield  {author} {\bibinfo {author} {\bibfnamefont {T.~G.}\ \bibnamefont
  {Sano}}\ and\ \bibinfo {author} {\bibfnamefont {H.}~\bibnamefont
  {Hayakawa}},\ }\href@noop {} {\bibfield  {journal} {\bibinfo  {journal}
  {Phys. Rev. E}\ }\textbf {\bibinfo {volume} {89}},\ \bibinfo {pages} {032104}
  (\bibinfo {year} {2014})}\BibitemShut {NoStop}%
\bibitem [{\citenamefont {Chen}\ and\ \citenamefont
  {Just}(2014)}]{chen2014large}%
  \BibitemOpen
  \bibfield  {author} {\bibinfo {author} {\bibfnamefont {Y.}~\bibnamefont
  {Chen}}\ and\ \bibinfo {author} {\bibfnamefont {W.}~\bibnamefont {Just}},\
  }\href@noop {} {\bibfield  {journal} {\bibinfo  {journal} {Phys. Rev. E}\
  }\textbf {\bibinfo {volume} {90}},\ \bibinfo {pages} {042102} (\bibinfo
  {year} {2014})}\BibitemShut {NoStop}%
\bibitem [{\citenamefont {Menzel}(2015)}]{menzel2015tuned}%
  \BibitemOpen
  \bibfield  {author} {\bibinfo {author} {\bibfnamefont {A.~M.}\ \bibnamefont
  {Menzel}},\ }\href@noop {} {\bibfield  {journal} {\bibinfo  {journal} {Phys.
  Rep.}\ }\textbf {\bibinfo {volume} {554}},\ \bibinfo {pages} {1} (\bibinfo
  {year} {2015})}\BibitemShut {NoStop}%
\bibitem [{\citenamefont {Fall}\ \emph {et~al.}(2008)\citenamefont {Fall},
  \citenamefont {Huang}, \citenamefont {Bertrand}, \citenamefont {Ovarlez},\
  and\ \citenamefont {Bonn}}]{fall2008shear}%
  \BibitemOpen
  \bibfield  {author} {\bibinfo {author} {\bibfnamefont {A.}~\bibnamefont
  {Fall}}, \bibinfo {author} {\bibfnamefont {N.}~\bibnamefont {Huang}},
  \bibinfo {author} {\bibfnamefont {F.}~\bibnamefont {Bertrand}}, \bibinfo
  {author} {\bibfnamefont {G.}~\bibnamefont {Ovarlez}}, \ and\ \bibinfo
  {author} {\bibfnamefont {D.}~\bibnamefont {Bonn}},\ }\href@noop {} {\bibfield
   {journal} {\bibinfo  {journal} {Phys. Rev. Lett.}\ }\textbf {\bibinfo
  {volume} {100}},\ \bibinfo {pages} {018301} (\bibinfo {year}
  {2008})}\BibitemShut {NoStop}%
\bibitem [{\citenamefont {Yamamoto}\ and\ \citenamefont
  {Schweizer}(2014)}]{yamamoto2014microscopic}%
  \BibitemOpen
  \bibfield  {author} {\bibinfo {author} {\bibfnamefont {U.}~\bibnamefont
  {Yamamoto}}\ and\ \bibinfo {author} {\bibfnamefont {K.~S.}\ \bibnamefont
  {Schweizer}},\ }\href@noop {} {\bibfield  {journal} {\bibinfo  {journal}
  {Macromolecules}\ }\textbf {\bibinfo {volume} {48}},\ \bibinfo {pages} {152}
  (\bibinfo {year} {2014})}\BibitemShut {NoStop}%
\bibitem [{\citenamefont {Torchilin}(2000)}]{torchilin2000drug}%
  \BibitemOpen
  \bibfield  {author} {\bibinfo {author} {\bibfnamefont {V.~P.}\ \bibnamefont
  {Torchilin}},\ }\href@noop {} {\bibfield  {journal} {\bibinfo  {journal}
  {Eur. J. Pharm. Sci.}\ }\textbf {\bibinfo {volume} {11}},\ \bibinfo {pages}
  {S81} (\bibinfo {year} {2000})}\BibitemShut {NoStop}%
\bibitem [{\citenamefont {Peer}\ \emph {et~al.}(2007)\citenamefont {Peer},
  \citenamefont {Karp}, \citenamefont {Hong}, \citenamefont {Farokhzad},
  \citenamefont {Margalit},\ and\ \citenamefont
  {Langer}}]{peer2007nanocarriers}%
  \BibitemOpen
  \bibfield  {author} {\bibinfo {author} {\bibfnamefont {D.}~\bibnamefont
  {Peer}}, \bibinfo {author} {\bibfnamefont {J.~M.}\ \bibnamefont {Karp}},
  \bibinfo {author} {\bibfnamefont {S.}~\bibnamefont {Hong}}, \bibinfo {author}
  {\bibfnamefont {O.~C.}\ \bibnamefont {Farokhzad}}, \bibinfo {author}
  {\bibfnamefont {R.}~\bibnamefont {Margalit}}, \ and\ \bibinfo {author}
  {\bibfnamefont {R.}~\bibnamefont {Langer}},\ }\href@noop {} {\bibfield
  {journal} {\bibinfo  {journal} {Nature Nanotech.}\ }\textbf {\bibinfo
  {volume} {2}},\ \bibinfo {pages} {751} (\bibinfo {year} {2007})}\BibitemShut
  {NoStop}%
\bibitem [{\citenamefont {Bertrand}\ \emph {et~al.}(2014)\citenamefont
  {Bertrand}, \citenamefont {Wu}, \citenamefont {Xu}, \citenamefont {Kamaly},\
  and\ \citenamefont {Farokhzad}}]{bertrand2014cancer}%
  \BibitemOpen
  \bibfield  {author} {\bibinfo {author} {\bibfnamefont {N.}~\bibnamefont
  {Bertrand}}, \bibinfo {author} {\bibfnamefont {J.}~\bibnamefont {Wu}},
  \bibinfo {author} {\bibfnamefont {X.}~\bibnamefont {Xu}}, \bibinfo {author}
  {\bibfnamefont {N.}~\bibnamefont {Kamaly}}, \ and\ \bibinfo {author}
  {\bibfnamefont {O.~C.}\ \bibnamefont {Farokhzad}},\ }\href@noop {} {\bibfield
   {journal} {\bibinfo  {journal} {Adv. Drug Delivery Rev.}\ }\textbf {\bibinfo
  {volume} {66}},\ \bibinfo {pages} {2} (\bibinfo {year} {2014})}\BibitemShut
  {NoStop}%
\bibitem [{\citenamefont {Alexiou}\ \emph {et~al.}(2006)\citenamefont
  {Alexiou}, \citenamefont {Schmid}, \citenamefont {Jurgons}, \citenamefont
  {Kremer}, \citenamefont {Wanner}, \citenamefont {Bergemann}, \citenamefont
  {Huenges}, \citenamefont {Nawroth}, \citenamefont {Arnold},\ and\
  \citenamefont {Parak}}]{alexiou2006targeting}%
  \BibitemOpen
  \bibfield  {author} {\bibinfo {author} {\bibfnamefont {C.}~\bibnamefont
  {Alexiou}}, \bibinfo {author} {\bibfnamefont {R.~J.}\ \bibnamefont {Schmid}},
  \bibinfo {author} {\bibfnamefont {R.}~\bibnamefont {Jurgons}}, \bibinfo
  {author} {\bibfnamefont {M.}~\bibnamefont {Kremer}}, \bibinfo {author}
  {\bibfnamefont {G.}~\bibnamefont {Wanner}}, \bibinfo {author} {\bibfnamefont
  {C.}~\bibnamefont {Bergemann}}, \bibinfo {author} {\bibfnamefont
  {E.}~\bibnamefont {Huenges}}, \bibinfo {author} {\bibfnamefont
  {T.}~\bibnamefont {Nawroth}}, \bibinfo {author} {\bibfnamefont
  {W.}~\bibnamefont {Arnold}}, \ and\ \bibinfo {author} {\bibfnamefont {F.~G.}\
  \bibnamefont {Parak}},\ }\href@noop {} {\bibfield  {journal} {\bibinfo
  {journal} {Eur. Biophys. J.}\ }\textbf {\bibinfo {volume} {35}},\ \bibinfo
  {pages} {446} (\bibinfo {year} {2006})}\BibitemShut {NoStop}%
\bibitem [{\citenamefont {Tietze}\ \emph {et~al.}(2013)\citenamefont {Tietze},
  \citenamefont {Lyer}, \citenamefont {D{\"u}rr}, \citenamefont {Struffert},
  \citenamefont {Engelhorn}, \citenamefont {Schwarz}, \citenamefont {Eckert},
  \citenamefont {G{\"o}en}, \citenamefont {Vasylyev}, \citenamefont {Peukert},
  \citenamefont {Wiekhorst}, \citenamefont {Trahms}, \citenamefont
  {D{\"o}rfler},\ and\ \citenamefont {Alexiou}}]{tietze2013efficient}%
  \BibitemOpen
  \bibfield  {author} {\bibinfo {author} {\bibfnamefont {R.}~\bibnamefont
  {Tietze}}, \bibinfo {author} {\bibfnamefont {S.}~\bibnamefont {Lyer}},
  \bibinfo {author} {\bibfnamefont {S.}~\bibnamefont {D{\"u}rr}}, \bibinfo
  {author} {\bibfnamefont {T.}~\bibnamefont {Struffert}}, \bibinfo {author}
  {\bibfnamefont {T.}~\bibnamefont {Engelhorn}}, \bibinfo {author}
  {\bibfnamefont {M.}~\bibnamefont {Schwarz}}, \bibinfo {author} {\bibfnamefont
  {E.}~\bibnamefont {Eckert}}, \bibinfo {author} {\bibfnamefont
  {T.}~\bibnamefont {G{\"o}en}}, \bibinfo {author} {\bibfnamefont
  {S.}~\bibnamefont {Vasylyev}}, \bibinfo {author} {\bibfnamefont
  {W.}~\bibnamefont {Peukert}}, \bibinfo {author} {\bibfnamefont
  {F.}~\bibnamefont {Wiekhorst}}, \bibinfo {author} {\bibfnamefont
  {L.}~\bibnamefont {Trahms}}, \bibinfo {author} {\bibfnamefont
  {A.}~\bibnamefont {D{\"o}rfler}}, \ and\ \bibinfo {author} {\bibfnamefont
  {C.}~\bibnamefont {Alexiou}},\ }\href@noop {} {\bibfield  {journal} {\bibinfo
   {journal} {Nanomed. Nanotechnol.}\ }\textbf {\bibinfo {volume} {9}},\
  \bibinfo {pages} {961} (\bibinfo {year} {2013})}\BibitemShut {NoStop}%
\bibitem [{\citenamefont {Zaloga}\ \emph {et~al.}(2014)\citenamefont {Zaloga},
  \citenamefont {Janko}, \citenamefont {Nowak}, \citenamefont {Matuszak},
  \citenamefont {Knaup}, \citenamefont {Eberbeck}, \citenamefont {Tietze},
  \citenamefont {Unterweger}, \citenamefont {Friedrich}, \citenamefont {Duerr},
  \citenamefont {Heimke-Brinck}, \citenamefont {Baum}, \citenamefont {Cicha},
  \citenamefont {D{\"o}rje}, \citenamefont {Odenbach}, \citenamefont {Lyer},
  \citenamefont {Lee},\ and\ \citenamefont {Alexiou}}]{zaloga2014development}%
  \BibitemOpen
  \bibfield  {author} {\bibinfo {author} {\bibfnamefont {J.}~\bibnamefont
  {Zaloga}}, \bibinfo {author} {\bibfnamefont {C.}~\bibnamefont {Janko}},
  \bibinfo {author} {\bibfnamefont {J.}~\bibnamefont {Nowak}}, \bibinfo
  {author} {\bibfnamefont {J.}~\bibnamefont {Matuszak}}, \bibinfo {author}
  {\bibfnamefont {S.}~\bibnamefont {Knaup}}, \bibinfo {author} {\bibfnamefont
  {D.}~\bibnamefont {Eberbeck}}, \bibinfo {author} {\bibfnamefont
  {R.}~\bibnamefont {Tietze}}, \bibinfo {author} {\bibfnamefont
  {H.}~\bibnamefont {Unterweger}}, \bibinfo {author} {\bibfnamefont {R.~P.}\
  \bibnamefont {Friedrich}}, \bibinfo {author} {\bibfnamefont {S.}~\bibnamefont
  {Duerr}}, \bibinfo {author} {\bibfnamefont {R.}~\bibnamefont
  {Heimke-Brinck}}, \bibinfo {author} {\bibfnamefont {E.}~\bibnamefont {Baum}},
  \bibinfo {author} {\bibfnamefont {I.}~\bibnamefont {Cicha}}, \bibinfo
  {author} {\bibfnamefont {F.}~\bibnamefont {D{\"o}rje}}, \bibinfo {author}
  {\bibfnamefont {S.}~\bibnamefont {Odenbach}}, \bibinfo {author}
  {\bibfnamefont {S.}~\bibnamefont {Lyer}}, \bibinfo {author} {\bibfnamefont
  {G.}~\bibnamefont {Lee}}, \ and\ \bibinfo {author} {\bibfnamefont
  {C.}~\bibnamefont {Alexiou}},\ }\href@noop {} {\bibfield  {journal} {\bibinfo
   {journal} {Int. J. Nanomedicine}\ }\textbf {\bibinfo {volume} {9}},\
  \bibinfo {pages} {4847} (\bibinfo {year} {2014})}\BibitemShut {NoStop}%
\bibitem [{\citenamefont {Dobson}(2006)}]{dobson2006magnetic}%
  \BibitemOpen
  \bibfield  {author} {\bibinfo {author} {\bibfnamefont {J.}~\bibnamefont
  {Dobson}},\ }\href@noop {} {\bibfield  {journal} {\bibinfo  {journal} {Drug
  Develop. Res.}\ }\textbf {\bibinfo {volume} {67}},\ \bibinfo {pages} {55}
  (\bibinfo {year} {2006})}\BibitemShut {NoStop}%
\bibitem [{\citenamefont {Matuszak}\ \emph {et~al.}(2015)\citenamefont
  {Matuszak}, \citenamefont {Zaloga}, \citenamefont {Friedrich}, \citenamefont
  {Lyer}, \citenamefont {Nowak}, \citenamefont {Odenbach}, \citenamefont
  {Alexiou},\ and\ \citenamefont {Cicha}}]{matuszak2015endothelial}%
  \BibitemOpen
  \bibfield  {author} {\bibinfo {author} {\bibfnamefont {J.}~\bibnamefont
  {Matuszak}}, \bibinfo {author} {\bibfnamefont {J.}~\bibnamefont {Zaloga}},
  \bibinfo {author} {\bibfnamefont {R.~P.}\ \bibnamefont {Friedrich}}, \bibinfo
  {author} {\bibfnamefont {S.}~\bibnamefont {Lyer}}, \bibinfo {author}
  {\bibfnamefont {J.}~\bibnamefont {Nowak}}, \bibinfo {author} {\bibfnamefont
  {S.}~\bibnamefont {Odenbach}}, \bibinfo {author} {\bibfnamefont
  {C.}~\bibnamefont {Alexiou}}, \ and\ \bibinfo {author} {\bibfnamefont
  {I.}~\bibnamefont {Cicha}},\ }\href@noop {} {\bibfield  {journal} {\bibinfo
  {journal} {J. Magn. Magn. Mater.}\ }\textbf {\bibinfo {volume} {380}},\
  \bibinfo {pages} {20} (\bibinfo {year} {2015})}\BibitemShut {NoStop}%
\bibitem [{\citenamefont {Dhont}(1996)}]{dhont1996introduction}%
  \BibitemOpen
  \bibfield  {author} {\bibinfo {author} {\bibfnamefont {J.~K.~G.}\
  \bibnamefont {Dhont}},\ }\href@noop {} {\emph {\bibinfo {title} {An
  Introduction to Dynamics of Colloids}}}\ (\bibinfo  {publisher} {Elsevier
  Science, Amsterdam},\ \bibinfo {year} {1996})\BibitemShut {NoStop}%
\bibitem [{\citenamefont {Risken}(1996)}]{risken1996fokker}%
  \BibitemOpen
  \bibfield  {author} {\bibinfo {author} {\bibfnamefont {H.}~\bibnamefont
  {Risken}},\ }\href@noop {} {\emph {\bibinfo {title} {{The Fokker-Planck
  Equation}}}}\ (\bibinfo  {publisher} {Springer, Berlin},\ \bibinfo {year}
  {1996})\BibitemShut {NoStop}%
\bibitem [{\citenamefont {Press}\ \emph {et~al.}(1992)\citenamefont {Press},
  \citenamefont {Teukolsky}, \citenamefont {Vetterling},\ and\ \citenamefont
  {Flannery}}]{press1992numerical}%
  \BibitemOpen
  \bibfield  {author} {\bibinfo {author} {\bibfnamefont {W.~H.}\ \bibnamefont
  {Press}}, \bibinfo {author} {\bibfnamefont {S.~A.}\ \bibnamefont
  {Teukolsky}}, \bibinfo {author} {\bibfnamefont {W.~T.}\ \bibnamefont
  {Vetterling}}, \ and\ \bibinfo {author} {\bibfnamefont {B.~P.}\ \bibnamefont
  {Flannery}},\ }\href@noop {} {\emph {\bibinfo {title} {Numerical Recipes in
  C}}}\ (\bibinfo  {publisher} {Cambridge University Press, Cambridge},\
  \bibinfo {year} {1992})\BibitemShut {NoStop}%
\bibitem [{\citenamefont {Landau}\ and\ \citenamefont
  {Lifshitz}(1977)}]{landau1077quantum}%
  \BibitemOpen
  \bibfield  {author} {\bibinfo {author} {\bibfnamefont {L.~D.}\ \bibnamefont
  {Landau}}\ and\ \bibinfo {author} {\bibfnamefont {E.~M.}\ \bibnamefont
  {Lifshitz}},\ }\href@noop {} {\emph {\bibinfo {title} {Quantum Mechanics
  (Non-relativistic Theory)}}}\ (\bibinfo  {publisher} {Pergamon Press,
  Oxford},\ \bibinfo {year} {1977})\BibitemShut {NoStop}%
\bibitem [{\citenamefont {Thurm}\ and\ \citenamefont
  {Odenbach}(2002)}]{thurm2002magnetic}%
  \BibitemOpen
  \bibfield  {author} {\bibinfo {author} {\bibfnamefont {S.}~\bibnamefont
  {Thurm}}\ and\ \bibinfo {author} {\bibfnamefont {S.}~\bibnamefont
  {Odenbach}},\ }\href@noop {} {\bibfield  {journal} {\bibinfo  {journal} {J.
  Magn. Magn. Mater.}\ }\textbf {\bibinfo {volume} {252}},\ \bibinfo {pages}
  {247} (\bibinfo {year} {2002})}\BibitemShut {NoStop}%
\bibitem [{\citenamefont {Odenbach}(2003)}]{odenbach2003ferrofluids}%
  \BibitemOpen
  \bibfield  {author} {\bibinfo {author} {\bibfnamefont {S.}~\bibnamefont
  {Odenbach}},\ }\href@noop {} {\bibfield  {journal} {\bibinfo  {journal}
  {Colloid Surface A}\ }\textbf {\bibinfo {volume} {217}},\ \bibinfo {pages}
  {171} (\bibinfo {year} {2003})}\BibitemShut {NoStop}%
\bibitem [{\citenamefont {Huke}\ and\ \citenamefont
  {L{\"u}cke}(2004)}]{huke2004magnetic}%
  \BibitemOpen
  \bibfield  {author} {\bibinfo {author} {\bibfnamefont {B.}~\bibnamefont
  {Huke}}\ and\ \bibinfo {author} {\bibfnamefont {M.}~\bibnamefont
  {L{\"u}cke}},\ }\href@noop {} {\bibfield  {journal} {\bibinfo  {journal}
  {Rep. Prog. Phys.}\ }\textbf {\bibinfo {volume} {67}},\ \bibinfo {pages}
  {1731} (\bibinfo {year} {2004})}\BibitemShut {NoStop}%
\bibitem [{\citenamefont {Holm}\ and\ \citenamefont
  {Weis}(2005)}]{holm2005structure}%
  \BibitemOpen
  \bibfield  {author} {\bibinfo {author} {\bibfnamefont {C.}~\bibnamefont
  {Holm}}\ and\ \bibinfo {author} {\bibfnamefont {J.-J.}\ \bibnamefont
  {Weis}},\ }\href@noop {} {\bibfield  {journal} {\bibinfo  {journal} {Curr.
  Opin. Colloid Interface Sci.}\ }\textbf {\bibinfo {volume} {10}},\ \bibinfo
  {pages} {133} (\bibinfo {year} {2005})}\BibitemShut {NoStop}%
\bibitem [{\citenamefont {Klapp}(2005)}]{klapp2005dipolar}%
  \BibitemOpen
  \bibfield  {author} {\bibinfo {author} {\bibfnamefont {S.~H.~L.}\
  \bibnamefont {Klapp}},\ }\href@noop {} {\bibfield  {journal} {\bibinfo
  {journal} {J. Phys.: Condens. Matter}\ }\textbf {\bibinfo {volume} {17}},\
  \bibinfo {pages} {R525} (\bibinfo {year} {2005})}\BibitemShut {NoStop}%
\bibitem [{\citenamefont {Mason}\ \emph {et~al.}(1997)\citenamefont {Mason},
  \citenamefont {Ganesan}, \citenamefont {Van~Zanten}, \citenamefont {Wirtz},\
  and\ \citenamefont {Kuo}}]{mason1997particle}%
  \BibitemOpen
  \bibfield  {author} {\bibinfo {author} {\bibfnamefont {T.~G.}\ \bibnamefont
  {Mason}}, \bibinfo {author} {\bibfnamefont {K.}~\bibnamefont {Ganesan}},
  \bibinfo {author} {\bibfnamefont {J.~H.}\ \bibnamefont {Van~Zanten}},
  \bibinfo {author} {\bibfnamefont {D.}~\bibnamefont {Wirtz}}, \ and\ \bibinfo
  {author} {\bibfnamefont {S.~C.}\ \bibnamefont {Kuo}},\ }\href@noop {}
  {\bibfield  {journal} {\bibinfo  {journal} {Phys. Rev. Lett.}\ }\textbf
  {\bibinfo {volume} {79}},\ \bibinfo {pages} {3282} (\bibinfo {year}
  {1997})}\BibitemShut {NoStop}%
\bibitem [{\citenamefont {Anthony}\ \emph {et~al.}(2006)\citenamefont
  {Anthony}, \citenamefont {Zhang},\ and\ \citenamefont
  {Granick}}]{anthony2006methods}%
  \BibitemOpen
  \bibfield  {author} {\bibinfo {author} {\bibfnamefont {S.}~\bibnamefont
  {Anthony}}, \bibinfo {author} {\bibfnamefont {L.}~\bibnamefont {Zhang}}, \
  and\ \bibinfo {author} {\bibfnamefont {S.}~\bibnamefont {Granick}},\
  }\href@noop {} {\bibfield  {journal} {\bibinfo  {journal} {Langmuir}\
  }\textbf {\bibinfo {volume} {22}},\ \bibinfo {pages} {5266} (\bibinfo {year}
  {2006})}\BibitemShut {NoStop}%
\bibitem [{\citenamefont {Saxton}(2008)}]{saxton2008single}%
  \BibitemOpen
  \bibfield  {author} {\bibinfo {author} {\bibfnamefont {M.~J.}\ \bibnamefont
  {Saxton}},\ }\href@noop {} {\bibfield  {journal} {\bibinfo  {journal} {Nature
  Methods}\ }\textbf {\bibinfo {volume} {5}},\ \bibinfo {pages} {671} (\bibinfo
  {year} {2008})}\BibitemShut {NoStop}%
\bibitem [{\citenamefont {Roth}\ \emph {et~al.}(2012)\citenamefont {Roth},
  \citenamefont {Schilde}, \citenamefont {Lellig}, \citenamefont {Kwade},\ and\
  \citenamefont {Auernhammer}}]{roth2012simultaneous}%
  \BibitemOpen
  \bibfield  {author} {\bibinfo {author} {\bibfnamefont {M.}~\bibnamefont
  {Roth}}, \bibinfo {author} {\bibfnamefont {C.}~\bibnamefont {Schilde}},
  \bibinfo {author} {\bibfnamefont {P.}~\bibnamefont {Lellig}}, \bibinfo
  {author} {\bibfnamefont {A.}~\bibnamefont {Kwade}}, \ and\ \bibinfo {author}
  {\bibfnamefont {G.~K.}\ \bibnamefont {Auernhammer}},\ }\href@noop {}
  {\bibfield  {journal} {\bibinfo  {journal} {Chem. Lett.}\ }\textbf {\bibinfo
  {volume} {41}},\ \bibinfo {pages} {1110} (\bibinfo {year}
  {2012})}\BibitemShut {NoStop}%
\bibitem [{\citenamefont {Huang}\ \emph {et~al.}(2015)\citenamefont {Huang},
  \citenamefont {Pessot}, \citenamefont {Cremer}, \citenamefont {Weeber},
  \citenamefont {Holm}, \citenamefont {Nowak}, \citenamefont {Odenbach},
  \citenamefont {Menzel},\ and\ \citenamefont
  {Auernhammer}}]{huang2015buckling}%
  \BibitemOpen
  \bibfield  {author} {\bibinfo {author} {\bibfnamefont {S.}~\bibnamefont
  {Huang}}, \bibinfo {author} {\bibfnamefont {G.}~\bibnamefont {Pessot}},
  \bibinfo {author} {\bibfnamefont {P.}~\bibnamefont {Cremer}}, \bibinfo
  {author} {\bibfnamefont {R.}~\bibnamefont {Weeber}}, \bibinfo {author}
  {\bibfnamefont {C.}~\bibnamefont {Holm}}, \bibinfo {author} {\bibfnamefont
  {J.}~\bibnamefont {Nowak}}, \bibinfo {author} {\bibfnamefont
  {S.}~\bibnamefont {Odenbach}}, \bibinfo {author} {\bibfnamefont {A.~M.}\
  \bibnamefont {Menzel}}, \ and\ \bibinfo {author} {\bibfnamefont {G.~K.}\
  \bibnamefont {Auernhammer}},\ }\href {\doibase 10.1039/C5SM01814E} {\bibfield
   {journal} {\bibinfo  {journal} {Soft Matter}\ } (\bibinfo {year} {2015}),\
  10.1039/C5SM01814E}\BibitemShut {NoStop}%
\bibitem [{\citenamefont {Brown}\ and\ \citenamefont
  {Jaeger}(2014)}]{brown2014shear}%
  \BibitemOpen
  \bibfield  {author} {\bibinfo {author} {\bibfnamefont {E.}~\bibnamefont
  {Brown}}\ and\ \bibinfo {author} {\bibfnamefont {H.~M.}\ \bibnamefont
  {Jaeger}},\ }\href@noop {} {\bibfield  {journal} {\bibinfo  {journal} {Rep.
  Prog. Phys.}\ }\textbf {\bibinfo {volume} {77}},\ \bibinfo {pages} {046602}
  (\bibinfo {year} {2014})}\BibitemShut {NoStop}%
\bibitem [{\citenamefont {Baule}\ \emph {et~al.}(2010)\citenamefont {Baule},
  \citenamefont {Cohen},\ and\ \citenamefont {Touchette}}]{baule2010path}%
  \BibitemOpen
  \bibfield  {author} {\bibinfo {author} {\bibfnamefont {A.}~\bibnamefont
  {Baule}}, \bibinfo {author} {\bibfnamefont {E.~G.~D.}\ \bibnamefont {Cohen}},
  \ and\ \bibinfo {author} {\bibfnamefont {H.}~\bibnamefont {Touchette}},\
  }\href@noop {} {\bibfield  {journal} {\bibinfo  {journal} {J. Phys. A: Math.
  Theor.}\ }\textbf {\bibinfo {volume} {43}},\ \bibinfo {pages} {025003}
  (\bibinfo {year} {2010})}\BibitemShut {NoStop}%
\bibitem [{\citenamefont {Chen}\ \emph {et~al.}(2013)\citenamefont {Chen},
  \citenamefont {Baule}, \citenamefont {Touchette},\ and\ \citenamefont
  {Just}}]{chen2013weak}%
  \BibitemOpen
  \bibfield  {author} {\bibinfo {author} {\bibfnamefont {Y.}~\bibnamefont
  {Chen}}, \bibinfo {author} {\bibfnamefont {A.}~\bibnamefont {Baule}},
  \bibinfo {author} {\bibfnamefont {H.}~\bibnamefont {Touchette}}, \ and\
  \bibinfo {author} {\bibfnamefont {W.}~\bibnamefont {Just}},\ }\href@noop {}
  {\bibfield  {journal} {\bibinfo  {journal} {Phys. Rev. E}\ }\textbf {\bibinfo
  {volume} {88}},\ \bibinfo {pages} {052103} (\bibinfo {year}
  {2013})}\BibitemShut {NoStop}%
\bibitem [{\citenamefont {Majumdar}\ and\ \citenamefont
  {Comtet}(2002)}]{majumdar2002local}%
  \BibitemOpen
  \bibfield  {author} {\bibinfo {author} {\bibfnamefont {S.~N.}\ \bibnamefont
  {Majumdar}}\ and\ \bibinfo {author} {\bibfnamefont {A.}~\bibnamefont
  {Comtet}},\ }\href@noop {} {\bibfield  {journal} {\bibinfo  {journal} {Phys.
  Rev. Lett.}\ }\textbf {\bibinfo {volume} {89}},\ \bibinfo {pages} {060601}
  (\bibinfo {year} {2002})}\BibitemShut {NoStop}%
\end{thebibliography}

%

\end{document}